\documentclass[11pt,a4paper]{article}
\pdfoutput=1
\usepackage{jcappub} 
\usepackage[T1]{fontenc}
\usepackage[english]{babel}
\usepackage{algorithm}
\usepackage{algorithmic}
\usepackage{xcolor}
\usepackage{mathdots}
\usepackage{subcaption}
\usepackage{layout}
\usepackage{printlen}
\usepackage{hyperref}
\usepackage{comment}
\usepackage[colorinlistoftodos]{todonotes}
\usepackage[normalem]{ulem}
\usepackage{soul}
\usepackage{tikz}
\usepackage{float}
\usepackage{booktabs}
\usepackage{graphicx}

\notoc

\title{(The) Wiggles going non-linear}

\author{Nathan Cohen}
\author{and Jan Hamann}
\affiliation{Sydney Consortium for Particle Physics and Cosmology, School of Physics, The University
of New South Wales, Sydney NSW 2052, Australia}

\emailAdd{nathan.cohen@unsw.edu.au}
\emailAdd{jan.hamann@unsw.edu.au}

\abstract{The simplest models of slow-roll inflation predict a featureless, nearly scale-invariant power spectrum of primordial curvature perturbations, consistent with current observations. However, in many non-minimal realisations of inflation, one generically expects the primordial power spectrum (PPS) to be ``wiggly'' with features that strongly deviate from scale-invariance. Current and next generation large scale structure (LSS) surveys will probe the PPS with unprecedented accuracy and therefore also increase sensitivity to power spectrum wiggles. However, accessing the information contained in these data will require an understanding of the behaviour of wiggly power spectra beyond the linear regime of structure formation. In this work, we use high-resolution $N$-body simulations to study the non-linear evolution of scenarios in which the PPS has superimposed oscillations and test whether the leading-order Gaussian damping form motivated by IR-resummed perturbation theory can be used as a compact one-parameter model of their late-time evolution. We test the relative improvement in constraints by implementing this modelling strategy in a likelihood analysis via Gaussian Process Regression (GPR) emulation.  Paying special attention to identifying our approach's domain of validity and quantifying as well as propagating uncertainties, we demonstrate that as long as the frequency of the PPS modulation is large enough, we are able to predict the matter power spectrum with sub-percent accuracy – thereby enabling us to search for inflationary wiggles in LSS data.}

\begin{document}
\begin{flushright}
\large \tt{CPPC-2026-03}
\end{flushright}

\maketitle
\flushbottom

\section{Introduction}

Current cosmological observations are consistent with the simplest slow-roll inflationary picture, in which primordial scalar perturbations are nearly scale-invariant, Gaussian, and adiabatic~\cite{Planck2020}. Nevertheless, a range of well motivated inflationary scenarios predict departures from this minimal slow-roll picture, including localised features, oscillations, or stronger scale-dependent structure – collectively, primordial ``wiggles'' – in the primordial power spectrum (PPS)~\cite{Adams_2001, chen2011, Ach_carro_2013, Chluba:2015bqa}. Observations of the cosmic microwave background (CMB) allow for a reconstruction of this spectrum, but only over a finite range of comoving wavenumbers. On large scales, the accessible modes are limited by the size of the observable Universe, making these measurements strongly affected by cosmic variance. On small scales, the sensitivity is restricted by photon diffusion, instrumental resolution, and astrophysical foregrounds~\cite{Planck2018}. The primary CMB anisotropies also observe primordial information through a two-dimensional projection of the three-dimensional perturbation field, limiting the number of independent modes and smoothing detailed scale-dependent structure~\cite{Chen:2016FeaturesLSS, Ferraro2022LSS}.

By contrast, large-scale structure (LSS) surveys probe the evolved matter density field in three dimensions, accessing a much larger number of Fourier modes within the survey volume and extending the range of scales over which primordial structure can be tested~\cite{Chen:2016FeaturesLSS, Ferraro2022LSS, Taylor_2018, EuclidEmulator2}. In principle, these observations therefore provide a powerful and complementary probe of primordial wiggles that are difficult to detect in the CMB alone~\cite{Beutler2019, Taylor_2018, EuclidEmulator2}. 

Surveys such as the Dark Energy Spectroscopic Instrument (DESI)~\cite{Aghamousa:2016zmz}, Euclid~\cite{laureijs2011euclid} and LSST~\cite{allisonlsst}, are poised to provide high-precision measurements of large-scale structure, reaching approximately percent-level accuracy over a wide range of scales and redshifts~\cite{Taylor_2018, EuclidEmulator2}. The impact of this improvement on primordial-feature constraints has already been quantified on large or quasi-linear scales by restricting the galaxy-clustering information to $k<0.1\,h\,\mathrm{Mpc}^{-1}$, where non-linear corrections are not expected to dominate~\citep{Ballardini2016}. Related forecasts have also explored the complementarity of CMB and LSS information for primordial wiggles on large scales~\citep{Huang2012, Chen2016, Ballardini2018}. To fully exploit the information content of these surveys, however, it is necessary to extend the analysis to smaller scales ($k \gtrsim 0.1 \: h/\mathrm{Mpc}$), where non-linear gravitational evolution becomes important.

In linear theory, the late-time matter power spectrum is related to the PPS by transfer functions that encode the evolution of each Fourier mode through radiation and matter domination~\citep{Bardeen1986, Eisenstein_1999}. On sufficiently large scales this gives a direct mapping between primordial structure and the late-time matter power spectrum. On smaller scales, gravitational evolution couples Fourier modes and transfers power between them, so the final power spectrum is no longer a simple linear response to the primordial perturbations~\citep{Bernardeau_2002}. For oscillatory primordial features, this mode coupling is especially important: large-scale displacements smear initially coherent structure and damp any wiggles imprinted in the evolved matter power spectrum~\citep{Crocce_2006, Crocce_2008, Matsubara2008}. 

A variety of analytic approaches extend the linear description into this regime~\citep{Bernardeau_2013}. Standard perturbation theory expands the density and velocity fields order by order, so that non-linear mode coupling appears as a hierarchy of higher-order corrections to the linear power spectrum~\citep{Bernardeau_2002}. Resummed approaches reorganise this expansion by identifying classes of higher-order terms whose cumulative effect is important, particularly those associated with coherent large-scale displacements, and incorporating their contribution more directly~\citep{Matsubara2008, Crocce_2006, Crocce_2008}. Effective-field-theory approaches further extend the perturbative description by adding symmetry-allowed counterterms that encode the impact of short-scale dynamics on large-scale observables~\citep{Carrasco_2012, Baumann_2012}. These methods are physically motivated and computationally efficient, but they only remain accurate enough within a limited range of redshifts and wavenumbers.

By contrast, $N$-body simulations directly evolve the gravitational dynamics of structure formation and provide the gold standard for non-linear evolution on small scales~\cite{Springel_2005, AnguloHahn2022}.  Accurate predictions from direct simulations are, however, computationally demanding. Percent-level precision on the absolute matter power spectrum requires large simulation volumes and high particle counts, making brute-force simulation impractical for the many parameter combinations required in cosmological inference~\citep{Wong2020}.

Simulation-calibrated models mitigate this cost by replacing repeated $N$-body calculations with fast-to-evaluate approximations. These include physically motivated fitting formulae such as \texttt{HALOFIT}~\citep{HaloFit2012}, or \texttt{HMCode}~\citep{HMCode2021}, and emulators that interpolate simulation outputs across wavenumber, redshift, and cosmological parameters~\cite{Heitmann_2009,Heitmann_2013,EuclidEmulator2, Lawrence_2017, Angulo_2021}. Modern emulators can reach percent-level accuracy across extended cosmological parameter spaces, including models with massive neutrinos and dynamical dark energy~\citep{Heitmann_2010, EuclidEmulator2, Lawrence_2017, Angulo_2021}. These methods, however, generally assume a smooth power-law primordial power spectrum.

For oscillatory primordial features, the problem can be recast around the feature contribution itself. Since the oscillations enter as a small modulation around a smooth reference spectrum, their non-linear evolution can be modelled separately from that of the smooth component. IR-resummed perturbation theory shows that coherent large-scale displacements produce a leading exponential damping of the oscillatory contribution~\citep{Blas2016, Beutler2019, Vasudevan2019}. Motivated by this leading result, \citet{Ballardini_2020} treat the damping scale as an effective parameter and calibrate a one-parameter Gaussian envelope directly against simulations. The practical question is whether this leading-order approximation remains accurate enough over the scales, feature parameters and redshifts relevant to LSS analyses. Working with the relative matter power spectrum further isolates these wiggles and mitigates numerical effects common to matched feature and reference simulations~\citep{Wong2020}.

In this work we test this strategy, considering the example of global logarithmic and linear oscillatory features in the PPS. These templates provide controlled realisations of primordial wiggles whose non-linear evolution can be isolated and calibrated. We generate paired dark-matter-only $N$-body simulations, with and without primordial features, over wavenumbers $0.05 \, h \, \mathrm{Mpc}^{-1} \leq k \leq 0.6 \, h \, \mathrm{Mpc}^{-1}$ and at redshifts $z \in \{ 0, 1, 2, 3, 4, 5 \}$. For each feature type and frequency, we calibrate the Gaussian damping model of~\citet{Ballardini_2020} against the simulated relative matter power spectrum. We then use residual diagnostics to determine where the semi-analytic description is accurate – and where it fails. The output is a calibrated function $\Sigma(\omega, z)$, with statistical uncertainties and an empirically defined domain of validity. We interpolate this calibrated damping function using Gaussian Process Regression, producing a continuous and uncertainty-aware damping emulator and use it in a Euclid-like spectroscopic information comparison to quantify how calibrated non-linear damping changes the recoverable information on primordial wiggles.

This paper is structured as follows. In Sec.~\ref{TheoryBackground} we introduce the primordial feature templates, the relative matter power spectrum, the semi-analytic damping model and the Gaussian Process Regression method used to construct the emulator. In Sec.~\ref{AnalysisPipeline} we describe the $N$-body simulations, the matched-pair variance estimate, and the calibration of the damping function. In Sec.~\ref{results} we discuss our calibration diagnostics, identify the domain of validity of the damping model, construct the damping emulator, and demonstrate its use in a Euclid-like information comparison. We summarise our results and present our conclusions in Sec.~\ref{conclusion}.

\section{Theoretical Background}
\label{TheoryBackground}
\subsection{Superimposed Oscillations on the Primordial Power Spectrum} 
\label{WigglesPPS}

In the absence of features, the primordial power spectrum of curvature perturbations is well-approximated by a smooth power-law,
\begin{align}
    \mathcal{P}_\mathcal{R}^0(k)= A_\mathrm{s}\left(\frac{k}{k_*} \right)^{n_\mathrm{s} -1},
    \label{eq:base_PPS}
\end{align}
where $k$ is the comoving wavenumber, $n_\mathrm{s}$ is the scalar spectral index and $A_\mathrm{s}$ is the primordial amplitude evaluated at the pivot scale $k_* = 0.05 \: \mathrm{Mpc}^{-1}$, consistent with the convention adopted in analyses of CMB data \citep{Planck2020}. This form captures the leading prediction of slow roll inflation, a nearly scale invariant spectrum, with small deviations governed by the slow evolution of the inflaton field and background expansion rate~\cite{Stewart_1993, Lyth_1999}.

A wide range of well-motivated inflationary scenarios predict departures from this smooth spectrum leading to scale dependent features in $\mathcal{P}_{\mathcal{R}}(k)$. As reviewed in detail by~\citet{Chluba:2015bqa}, transient or periodic departures from the standard slow roll background, such as brief violations of slow roll or oscillatory dynamics of the inflationary background, can imprint characteristic oscillatory modulations in the primordial power spectrum. In many concrete realisations, these modulations are well approximated by corrections that are periodic in either the wavenumber itself, or its logarithm, depending on the underlying physical mechanism~\cite{Starobinsky1992, Adams_2001, Chen_2008}.

Rather than modelling specific inflationary models, we adopt the phenomenological oscillatory templates commonly employed in searches for primordial features in the CMB and LSS data~\cite{Planck2020, Meerburg_2014}, namely the linear oscillation model,
\begin{align}\label{eq:lin}
      \mathcal{P}^{\mathrm{lin}}_{\mathcal{R}^\mathrm{}}(k, \mathcal{A}, \omega, \phi) =\mathcal{P}_\mathcal{R}^0(k) \left(1 + \mathcal{A} \cos\left[\omega \, \frac{k}{k_*} + \phi \right] \right),
\end{align}
and the logarithmic oscillation model,
\begin{align}\label{eq:log}
      \mathcal{P}^{\mathrm{log}}_{\mathcal{R}^\mathrm{}}(k, \mathcal{A}, \omega, \phi) =\mathcal{P}_\mathcal{R}^0(k) \left(1 + \mathcal{A} \cos\left[\omega \, \log\frac{k}{k_*} + \phi \right] \right).
\end{align}
Here, $\mathcal{A}$ denotes the dimensionless modulation amplitude, $\omega$ is the oscillation frequency parameter and $\phi$ is a constant phase.

\subsection{Relative Power Spectra and Simulation Noise}
\label{UncPPS}

For a theoretical prediction of the PPS to be useful in a likelihood analysis, the modelling error of the corresponding observable should be subdominant to the measurement precision.  Since exploiting the full constraining power in current and next generation LSS measurements requires sub-percent predictions for the matter power spectrum~\cite{Taylor_2018, EuclidEmulator2}, we adopt a conservative accuracy target
\begin{align}
    \epsilon_0 \equiv \frac{\Delta\mathcal{P}(k)}{\mathcal{P}(k)} = 5 \times10^{-3}.
\end{align}
This threshold is used throughout the calibration to decide where the semi-analytic damping template is sufficiently accurate.

Achieving this accuracy requires tight control of both stochastic simulation noise and numerical systematics. If one were to try and directly predict the absolute matter power spectrum, these effects could easily obscure the small oscillatory features we wish to trace.

We therefore work with the relative matter power spectrum instead, defined as
\begin{align}\label{eq:relpowspec}
    \delta(k, \mathcal{A}, \omega, \phi, z) \equiv \frac{\mathcal{P}_\mathrm{m}(k, \mathcal{A}, \omega, \phi, z) - \mathcal{P}_\mathrm{m}^0(k,z)}{\mathcal{P}_\mathrm{m}^0(k,z)},
\end{align}
where $\mathcal{P}_\mathrm{m}(k, \mathcal{A}, \omega, \phi, z)$ is the matter power spectrum for the feature model and $\mathcal{P}_\mathrm{m}^0(k,z)$ is the corresponding featureless reference spectrum.

In the case of matched simulations, i.e., if the feature and reference simulations are generated with the same random seed (such that the two simulations share the same random Fourier phases in the initial density realisation), and provided the feature modulation is small enough so as not to affect the non-linear evolution in a radically different way, the fluctuations of the absolute power spectra due to sample variance can be approximately described by the same $k$-dependent function – which cancels out in the ratio.  Other numerical systematics due to, e.g., force softening, or time-integration errors likewise display a multiplicative behaviour and their influence is therefore similarly mitigated in the relative power spectrum~\cite{Wong2020}.

The usefulness of this quantity is further illustrated in linear theory. The mapping between the primordial curvature spectrum and the late-time matter power spectrum can be written as
\begin{equation}
    \mathcal{P}_\mathrm{m}(k,z) = \mathcal{T}^2(k,z) \, \mathcal{P}_\mathcal{R}^\mathrm{}(k),
\end{equation}
where the linear transfer function $\mathcal{T}^2(k,z)$ is independent of the detailed feature modulation. For a primordial spectrum of the form
\begin{align}
    \mathcal{P}_\mathcal{R}^\mathrm{}(k, \mathcal{A}, \omega, \phi) &=\mathcal{P}_\mathcal{R}^0(k) \left(1 + \mathcal{A} \cos\left[\omega f(k) + \phi \right] \right),
    \label{eq:smooth_and_osc_pps}
\end{align}
the transfer function cancels in the relative spectrum, giving
\begin{align}
    \delta_\mathrm{lin}(k, \mathcal{A}, \omega, \phi) &\equiv \mathcal{A} \cos\left[\omega f(k) + \phi \right].
    \label{eq:damp_ansatz}
\end{align}
Thus, before non-linear evolution is included, the relative matter power spectrum directly isolates the oscillatory feature contribution, and is independent of redshift too. Non-linear gravitational evolution then modifies this simple form, and introduces an additional dependence on $k$, as well as on redshift. The purpose of the damping model introduced in the next subsection is to accurately model that modification\footnote{In practice, a complete likelihood analysis would require a prediction for the smooth component of the non-linear matter power spectrum. This could be supplied by one of the simulation-calibrated fitting functions or emulators discussed earlier~\citep{HaloFit2012, HMCode2021,Heitmann_2013, EuclidEmulator2, Lawrence_2017, Angulo_2021}, with the calibrated damping envelope used to model the additional suppression of the primordial feature contribution.}.

\subsection{Semi-Analytic Damping Model}
\label{FittingTool}
On the quasi-linear to mildly non-linear scales targeted in this work, linear theory cannot provide a sufficiently accurate prediction of the late-time matter power spectrum. In this regime, non-linear gravitational evolution couples Fourier modes, redistributes power across scales, thereby distorting the imprints of primordial oscillations. The linear-theory relative spectrum in Eq.~\ref{eq:damp_ansatz} therefore provides only the undamped reference signal. 

We model this departure from linear theory using a semi-analytic damping template, following~\citet{Ballardini_2020}. Rather than predicting the full non-linear matter power spectrum, the template describes the suppression of the oscillatory feature contribution in the relative matter power spectrum. 
Our ansatz for the relative non-linear matter power spectrum is
\begin{align}
    \delta_{\text{fit}}(k, \mathcal{A}, \omega, \phi, z) &\equiv \mathcal{A} \cos\left[\omega f(k) + \phi \right] \; \mathcal{D}(k, \Sigma(\omega, z)),
    \label{nlmpsfit}
\end{align}
where $\mathcal{D}(k, \Sigma(\omega, z))$ captures the damping induced by non-linear evolution and $\Sigma(\omega, z)$ is an effective damping function to be calibrated from simulations.

The form of $\mathcal{D}$ is motivated by the well-studied damping of baryon acoustic oscillations (BAO). For BAO, the linear matter power spectrum can be decomposed into a smooth component and an oscillatory modulation in the same way as Eq.~\ref{eq:smooth_and_osc_pps}, 
\begin{align}
    \mathcal{P}_{\mathrm{m, lin}}(k, z) = \mathcal{P}_{\mathrm{m}, 0}(k, z)\left[ 1 + O(k)\right].
\end{align} 
Large-scale bulk motions suppress the oscillatory component of the power spectrum by displacing matter from its initial positions. In Lagrangian perturbation theory, the evolved position of a matter element is written as
\begin{align}
    \mathbf{x} = \mathbf{q} + \mathbf{\Psi}(\mathbf{q}, z),
\end{align}
where $\mathbf{q}$ is the initial Lagrangian coordinate and $\mathbf{\Psi}$ is the displacement field. The displacements smear initially coherent Fourier structure. At leading order their effect on the oscillatory component can be described by an approximately Gaussian damping whose scale is controlled by the variance of the displacement field~\cite{Crocce_2008, Matsubara2008, Eisenstein_2007}. The oscillatory component is therefore damped as
\begin{align}
    O(k) \rightarrow O(k)\exp\left[-\frac{k^2\Sigma^2(z)}{2}\right].
\end{align}

This displacement-driven suppression is familiar from BAO, for which the oscillatory component is accurately described by a Gaussian damping envelope~\citep{Eisenstein_2007, Seo2007}. The same leading exponential structure follows directly for primordial oscillatory features from IR-resummed perturbation theory~\citep{Blas2016, Beutler2019, Vasudevan2019}. These calculations resum the effect of coherent long-wavelength displacements on the wiggly component of the power spectrum. For linearly spaced oscillations, the leading result takes a BAO-like Gaussian form with a damping scale that depends only weakly on frequency. For logarithmically spaced oscillations, the complete leading result is generally more involved; the effective damping can depend explicitly on $k$, and an additional phase-mixing contribution can appear. Higher-order treatments add the remaining perturbative corrections and mixed contributions between BAO and primordial oscillations~\citep{Ballardini2024}.

Here we retain the leading Gaussian form
\begin{align}
    \mathcal{D}(k,\Sigma(\omega, z)) = \exp\left[\frac{-k^2 \Sigma^2(\omega, z)}{2}\right],
\end{align}
but treat $\Sigma(\omega, z)$ as an effective, $k$-independent damping scale fitted to the simulations. The model should therefore be understood as a one-parameter compression of the leading displacement-driven damping rather than as an evaluation of the complete IR-resummed expression. Its usefulness depends on whether the omitted scale dependence and additional perturbative contributions remain smaller than the accuracy required for an LSS analysis.

With this choice, the fitting template for the relative non-linear matter power spectrum becomes
\begin{align}
    \delta_{\text{fit}}(k, \mathcal{A}, \omega, \phi, z) &\equiv \mathcal{A} \cos\left[\omega f(k) + \phi \right]\exp\left[-\frac{k^2 \Sigma^2(\omega, z)}{2}\right].
    \label{eq:damping_template}
\end{align}
Here we allow $\Sigma(\omega, z)$ to depend on the feature frequency $\omega$, but it is held constant with respect to $k$ at each frequency and redshift. For linear oscillations, this closely follows the structure of the leading IR-resummed result. For logarithmic oscillations, it constitutes an additional approximation whose accuracy must be established against the simulations.

Equation~\ref{eq:damping_template} is the semi-analytic model used throughout this work: $\Sigma(\omega, z)$ is calibrated to the simulated relative power spectrum at different feature frequencies and redshifts. The calibration procedure and uncertainty estimate are described in Sec.~\ref{AnalysisPipeline} and interpolation of the resulting $\Sigma(\omega, z)$ in Sec.~\ref{GPRResults}. The damping envelope $\mathcal{D}(k, \Sigma(\omega, z))$ then captures the suppression of the oscillatory feature contribution and can be applied directly to the relative matter power spectrum.

\subsection{Gaussian Process Regression}
\label{GPR}
The simulation calibration provides $\Sigma(\omega, z)$ only at the sampled frequencies and redshifts. A likelihood analysis, however, requires the $\Sigma(\omega, z)$ to be evaluated continuously as the feature parameters are varied. We therefore use Gaussian Process Regression (GPR) to interpolate the calibrated damping function.

A Gaussian process (GP) is well suited to this task. The likelihood requires a continuous, uncertainty-aware prediction for $\Sigma(\omega, z)$, while the calibration itself does not prescribe a parametric form for its redshift and frequency dependence. A GP provides a smooth, non-parametric interpolation of the calibrated damping function, together with a predictive uncertainty that can be propagated into the likelihood calculation.

A GP defines a distribution over functions,
\begin{align}
    f(x) \sim \mathcal{GP}(m(x), K(x, x')),
\end{align}
where $m(x)$ is the mean function and $K(x, x')$ is the covariance function~\cite{rasmussen_williams_2006}. In this work the input is the feature frequency and redshift, $\mathbf{x}=(\omega, z)$, the target is the calibrated damping function, $f(\mathbf{x}) = \Sigma(\omega, z)$, and the uncertainty in each calibration fit, $\Delta\Sigma(\omega, z)$, is included as training noise.

As the damping function is expected to vary smoothly with frequency and redshift we use a radial basis function kernel,
\begin{align}
    K(\mathbf{x}, \mathbf{x}^{\prime}) = \sigma^2_s \exp\left[-\frac{1}{2}\sum_d\left(\frac{x_d - x_d^{\prime}}{\ell_d}\right)^2\right], 
\end{align}
which works well for smooth functions. For the training data, this covariance is augmented as
\begin{align}
    K_{ij}^{\mathrm{train}} = K(\mathbf{x}_i, \mathbf{x}_j) + [\Delta \Sigma(\mathbf{x}_i)]^2\delta_{ij}.
\end{align}
Here $\sigma_s$ is the kernel amplitude (or outputscale) which sets the typical variance of the prediction away from the training data. The quantities $\ell_d$ are the correlation lengths (or lengthscales) for each input dimension, and control how rapidly the prediction is allowed to vary along each direction. A large $\ell_d$ corresponds to smooth variation over that coordinate, while a small $\ell_d$ allows more rapid changes. Having a separate lengthscale for each dimension allows the GP to learn different characteristic variation scales in $\omega$ and $z$. The outputscale and lengthscales are known as the kernel hyperparameters and are determined from the data by maximising the marginal log-likelihood. 

In practice, we implement this emulation using the \texttt{scikit-learn}~\cite{scikit-learn} Gaussian process regressor. Once trained, evaluating the GP is numerically inexpensive, particularly when compared with generating a new matter power spectrum, making the emulator suitable for application to inference or model selection tasks.

\section{Simulations, Calibration and Accuracy Estimation}
\label{AnalysisPipeline}
\subsection{$N$-Body Simulation Initial Conditions}
\label{NBodyIC}

For each feature model and parameter combination listed in Tables~\ref{tab: $N$-bodyInit 1} and \ref{tab: $N$-bodyInit 2}, we generate a matched pair of primordial spectra, one with superimposed oscillatory features and one featureless reference spectrum. Each spectrum is evolved using \texttt{CLASS}~\cite{lesgourgues2011cosmic} to obtain the linear matter power spectrum at $z=65$, which is then passed to \texttt{N-GenIC}~\cite{VolkerNGenIC} to generate the particle initial conditions for the $N$-body simulations. 

For each feature-reference pair, the same random seed is used in \texttt{N-GenIC}, ensuring that the two simulations share the same initial random Fourier phases. Different matched pairs are generated with independent seeds\footnote{In principle, a single featureless reference simulation could be reused for multiple, or all, feature simulations. We instead generate a reference spectrum for each feature realisation so that the calibration samples the finite-realisation scatter of the relative spectrum itself, rather than conditioning all ratios on one particular reference density field.}. As discussed in Sec.~\ref{UncPPS}, this matched-pair construction suppresses sample variance and mitigates other numerical systematics that are approximately common to the feature and reference simulations when forming the relative matter power spectrum.

Each simulation uses $1,024^3$ particles in a comoving box of side length $1,024 \: \mathrm{Mpc}/h$. The initial redshift is $z = 65$ with snapshots saved at redshifts $z \in \{0, 1, 2, 3, 4, 5\}$. Particle evolution is performed using \texttt{Gadget-2}~\cite{Volkel2005}, and the evolved particle distributions are analysed using \texttt{GenPK}~\cite{2017ascl.soft06006B} to estimate each snapshot's matter power spectrum.

The calibration is performed over 
\begin{align}
    0.05 \:h\:\mathrm{Mpc}^{-1} \leq k \leq 0.6\:h\:\mathrm{Mpc}^{-1}.
\end{align}
This targets the weakly to moderately non-linear regime, where linear perturbation theory begins to break down but the imprint of primordial oscillatory features is still expected to be describable by a smooth, scale-dependent damping envelope~\cite{Ballardini_2020, Bernardeau_2002}.

The feature grid is chosen to cover a representative range of frequencies, phases, and redshifts while remaining numerically tractable. We fix the modulation amplitude to $\mathcal{A} =0.03$. Current CMB measurements constrain oscillatory features in the PPS to have relative amplitudes of at most a few percent, depending on the specific feature model and scale~\cite{Planck2020}. The adopted value is therefore broadly consistent with existing constraints while keeping the feature signal large enough to be robustly resolved in the simulations.

We consider $0\leq z \leq 5$, covering the epochs relevant to LSS surveys and higher-redshift probes of primordial information in the matter power spectrum. This includes spectroscopic galaxy surveys at $z \lesssim 2$ as well as probes such as intensity mapping and Lyman-$\alpha$ measurements extending to $z \gtrsim 4$~\cite{Aghamousa:2016zmz, laureijs2011euclid, allisonlsst}.

The oscillation frequency is varied to probe both slowly and rapidly varying primordial modulations within the fitted wavenumber interval. The central part of this range follows~\citet{Ballardini_2020}, with additional frequencies on either side to test the behaviour of the calibrated damping function beyond their grid. This choice is consistent with the frequency ranges explored in phenomenological searches for primordial features~\cite{Planck2020}. Very low frequencies vary slowly across the observable range and can become degenerate with smooth deformations of the primordial spectrum, such as a running of the spectral index~\cite{Planck2020}. Very high frequencies require finer wavenumber sampling and higher-resolution simulations to resolve the oscillatory structure. We therefore use a range broad enough to test the damping model, but restricted to features that remain distinguishable from smooth spectral deformations and numerically resolvable in the simulation outputs.

\begin{table}
\centering
\begin{subtable}{0.45\textwidth}
\centering
\begin{tabular}{|c|c|} \hline
 log$_{10}\omega$ & $\phi/\pi$ \\ \hline
 0.4 & 0.0 \\
 0.4 & 0.5 \\
 0.4 & 0.75 \\
 0.4 & 1 \\
 0.8 & 0.0 \\
 0.87 & 0.0 \\
 1.26 & 0.0 \\
 1.26 & 0.5 \\
 1.26 & 0.75 \\
 1.26 & 1 \\
 1.5 & 0.0 \\
 2.0 & 0.0 \\ \hline
\end{tabular}
\caption{Logarithmic feature-template parameters}
\label{tab: $N$-bodyInit 1}
\end{subtable}
\begin{subtable}{0.45\textwidth}
\centering
\begin{tabular}{|c|c|} \hline
 log$_{10}\omega$ & $\phi/\pi$ \\ \hline
 0.4 & 0.0 \\
 0.4 & 0.5 \\
 0.4 & 0.75 \\
 0.4 & 1 \\
 0.8 & 0.0 \\
 0.87 & 0.0 \\
 0.87 & 0.5 \\
 0.87 & 0.75 \\
 0.87 & 1 \\
 1.0 & 0.0 \\
 1.2 & 0.0 \\ \hline
\end{tabular}
\caption{Linear feature-template parameters}
\label{tab: $N$-bodyInit 2}
\end{subtable}
\caption{Feature configurations used for the $N$-body calibration. All simulations use $\mathcal{A} = 0.03$}
\end{table}
 
\subsection{$N$-Body Simulation Variance}
\label{SimVar}
In addition to the matched feature-reference simulation pairs used to sample the feature parameter space, we run four additional matched pairs with independent random seeds to estimate the variance in the relative non-linear matter power spectrum\footnote{The additional feature simulations use the logarithmic model with $\omega=0.4$, $\mathcal{A}=0.03$, $\phi = 0$.}. 

We compute the relative matter power spectrum for these four matched pairs and measure the sample variance across realisations at each wavenumber and redshift. The calibration likelihood uses a diagonal covariance, and we correct for the finite-sample bias introduced by estimating the covariance from a small number of simulations using the Hartlap correction~\citep{Hartlap_2006},
\begin{align}
     C_{ij}^{-1} = \frac{N_{\mathrm{sim}} - p - 2}{N_{\mathrm{sim}} - 2} \hat{C}_{ij}^{-1} \: \text{for} \: p < 2 N_{\mathrm{sim}} - 2,
\end{align}
where $N_{\mathrm{sim}}$ is the number of realisations and $p$ is the dimension of the covariance estimate.

Given our small number of realisations, the variance estimate will obviously be very noisy in $k$.  Before using it in the calibration likelihood as $\sigma_{\mathrm{sim}}(k, z)$, we therefore smooth $\sigma_{\mathrm{corr}}(k, z)$ with a polynomial fit. This smoothing preserves the broad $k$-dependence of the variance, but prevents individual noisy $k$-values from receiving artificially high or low inverse-variance weight in the calibration likelihood.

\begin{figure}[h]
    \centering
    \includegraphics[scale=0.75]{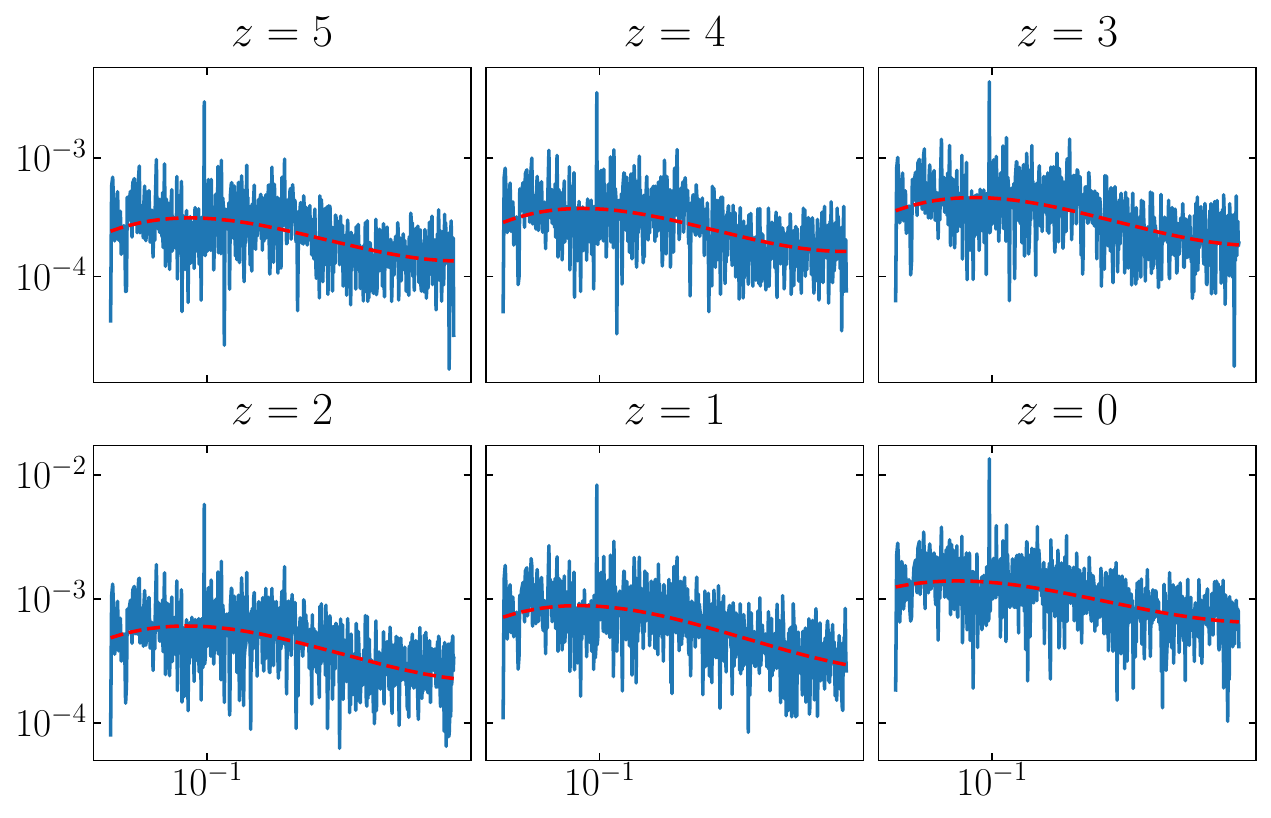}
    \caption{Polynomial smoothing of the corrected simulation uncertainty used in the calibration likelihood. The blue curve in each panel shows the corrected standard deviation estimated from the matched simulation pairs, while the red dashed curves show the smoothed approximation used for $\sigma_{\mathrm{sim}}(k,z)$.}
    \label{fig:VarianceApproximation}
\end{figure}

The resulting corrected relative uncertainty satisfies 
\begin{align}
    \frac{\sigma_{\mathrm{sim}}(k, z)}{|\delta_{\mathrm{sim}}(k, z)|} \lesssim 10^{-3}
\end{align} 
across the full wavenumber and redshift range considered, as shown in Fig.~\ref{fig:VarianceApproximation}. The stochastic simulation uncertainty is therefore at the per-mille level relative to the feature signal, well below the percent-level modulations considered here. This justifies using a single matched feature-reference pair at each sampled point in feature-parameter space for the main calibration.

\subsection{Model Calibration}
\label{ModelCal}

For each feature template, frequency, phase, and redshift, we calibrate $\Sigma(\omega, z)$ by fitting the semi-analytic damping template 
\begin{align}
    \delta_{\mathrm{fit}}(k, \mathcal{A}, \omega, \phi, z) = \mathcal{A}\cos[\omega f(k) + \phi]\exp\left[-\frac{k^2\Sigma^2(\omega, z)}{2}\right]
\end{align}
to the simulated relative matter power spectrum. The best-fitting value $\hat{\Sigma}(\omega, z)$ is obtained by minimising
\begin{align}
    \chi^{2}(\Sigma)=\sum_i\left[\frac{\delta_{\mathrm{fit}}(k_i, \Sigma)-\delta_{\mathrm{sim}}(k_i, z)}{\sigma_\mathrm{sim}(k_i, z)}\right]^{2},
    \label{eq:calibration_chisq}
\end{align}
over the fitted range $0.05 \leq k \leq 0.6 \: h \, \mathrm{Mpc}^{-1}$. Here $\mathcal{A}$, $\omega$, $\phi$, and $z$ are fixed for each calibration, so the only fitted parameter is $\Sigma$. The fit is performed independently for each sampled $(\omega, z, \phi)$ configuration using the \texttt{SciPy} least-squares optimiser \cite{2020SciPy-NMeth}. The statistical uncertainty on $\Sigma$ is obtained from the $\Delta\chi^2=1$ interval, 
\begin{align}
    \chi^2(\Sigma) = \chi^2(\hat{\Sigma}) + 1
    \label{eq:chisquared_unc}
\end{align}
which corresponds to the $1\sigma$ confidence interval for a single fitted parameter. This is the uncertainty propagated into the Gaussian process interpolation as described in Sec.~\ref{GPR}.

\section{Results}
\label{results}
\subsection{Calibration Diagnostics}

The calibration procedure has two purposes: to determine the best-fitting damping function and to identify the domain in which the one-parameter damping template is accurate enough for inference.  We therefore assess the model using two complementary diagnostics. First, $\Delta \Sigma(\omega, z)$ measures how precisely the damping function is constrained within the semi-analytic template, as described in Sec.~\ref{ModelCal}. Second, the residual diagnostics test whether the resulting damping template, $\delta_{\mathrm{fit}}$, reproduces the simulated relative matter power spectra to the required accuracy, $\epsilon_0$. A small $\Delta \Sigma(\omega, z)$ shows that the preferred damping function is well determined, but it does not by itself guarantee that the template is accurate across the fitted $k$-range.

To quantify the residual mismatch, we consider two measures: the maximum absolute residual
\begin{align}
    R_{\mathrm{max}} = \max_i \left|\delta_{\mathrm{fit}}(k_i) - \delta_{\mathrm{sim}}(k_i)\right|
    \label{eq:maximal_res}
\end{align}
and the global RMS residual
\begin{align}
    \sigma_\mathrm{abs} = \sqrt{\frac{1}{N_k}\sum_i \left[\delta_\mathrm{fit}(k_i) - \delta_\mathrm{sim}(k_i)\right]^2}.
    \label{RMS_res}
\end{align}
The former gives a conservative worst-case measure, while the latter measures the typical mismatch across the fitted range. We evaluate these both globally and in broad $k$-bins to track the scale-dependence of the residuals. Four equal sized bins are used in the latter case with $\Delta k \simeq 0.14 \, h\:\mathrm{Mpc}^{-1}$.

Finally, we define the fraction of Fourier modes exceeding the precision threshold,
\begin{align}
    f_{\mathrm{fail}}(\epsilon_0) = \frac{1}{N_k} \sum_{i=1}^{N_k}
    \begin{cases}
        1, & 
        \text{if } 
        |\delta_{\mathrm{fit}}(k_i) - \delta_{\mathrm{sim}}(k_i)| > \epsilon_0, 
        \\[12pt] 0, & \text{otherwise.}
    \end{cases}
\end{align}
This same accuracy scale is used in Sec.~\ref{ForecastResults} as a conservative model-error floor, linking the simulation validation directly to the inference comparison.

\subsection{Logarithmic Oscillations}
\label{LogResults}

\begin{figure}[h]
    \centering
    \includegraphics[width=\linewidth]{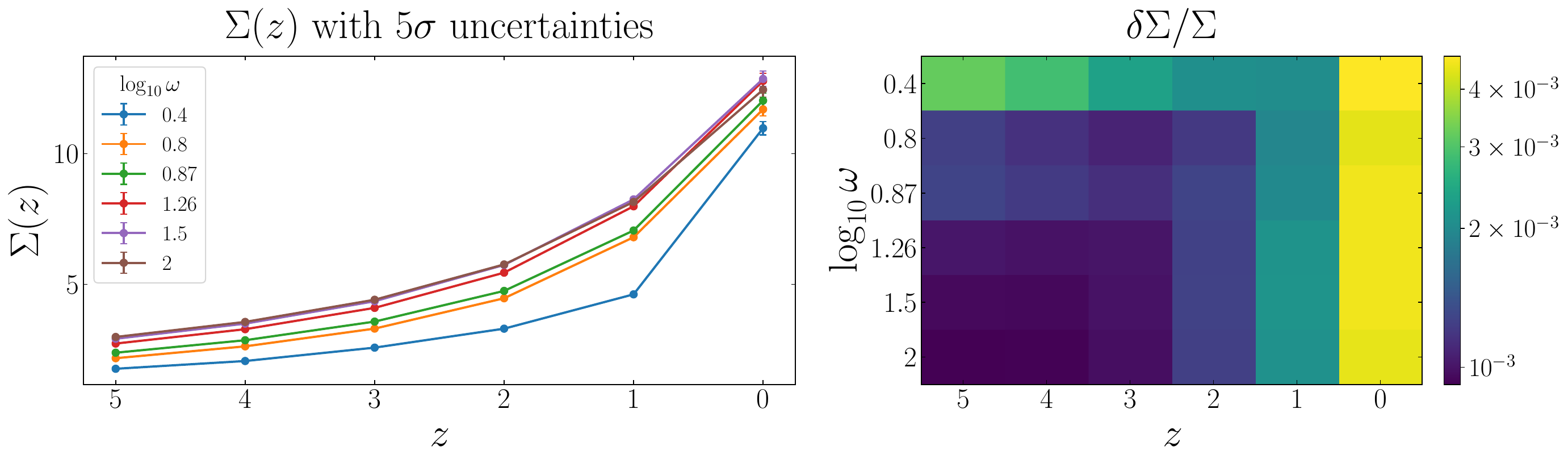}
    \caption{Fit diagnostics for the logarithmic frequency sweep. Left: calibrated $\Sigma(z)$ values. Right: relative uncertainty in the fitted damping function.}
    \label{fig:LogOsc_sigma_uncertianty_pair}
\end{figure}

We first apply the diagnostics to the logarithmic oscillations model. Calibrated damping functions and their relative uncertainties are shown in Fig.~\ref{fig:LogOsc_sigma_uncertianty_pair}. Across the sampled frequency-redshift grid, $\Delta \Sigma / \Sigma \leq 4\times10^{-3}$, indicating that the fitted $\Sigma$ values are well constrained. 

The fitted values of $\Sigma$ in Fig.~\ref{fig:LogOsc_sigma_uncertianty_pair} are consistent with its interpretation as an effective displacement scale controlling the suppression of coherent oscillatory power. At fixed frequency, $\Sigma$ increases towards lower redshift because the density field has undergone non-linear evolution for longer, and consequently mode coupling affects wider windows in $k$.

At fixed redshift, $\Sigma$ generally increases with oscillation frequency. Higher-frequency features have peaks and troughs that are closer together in $k$, so relatively local mode coupling can blur the oscillatory pattern. The peaks and troughs of lower-frequency features are further apart, and hence require stronger non-linear evolution before the oscillatory pattern is substantially smeared.

\begin{figure}[h]
    \centering
    \includegraphics[width=\linewidth]{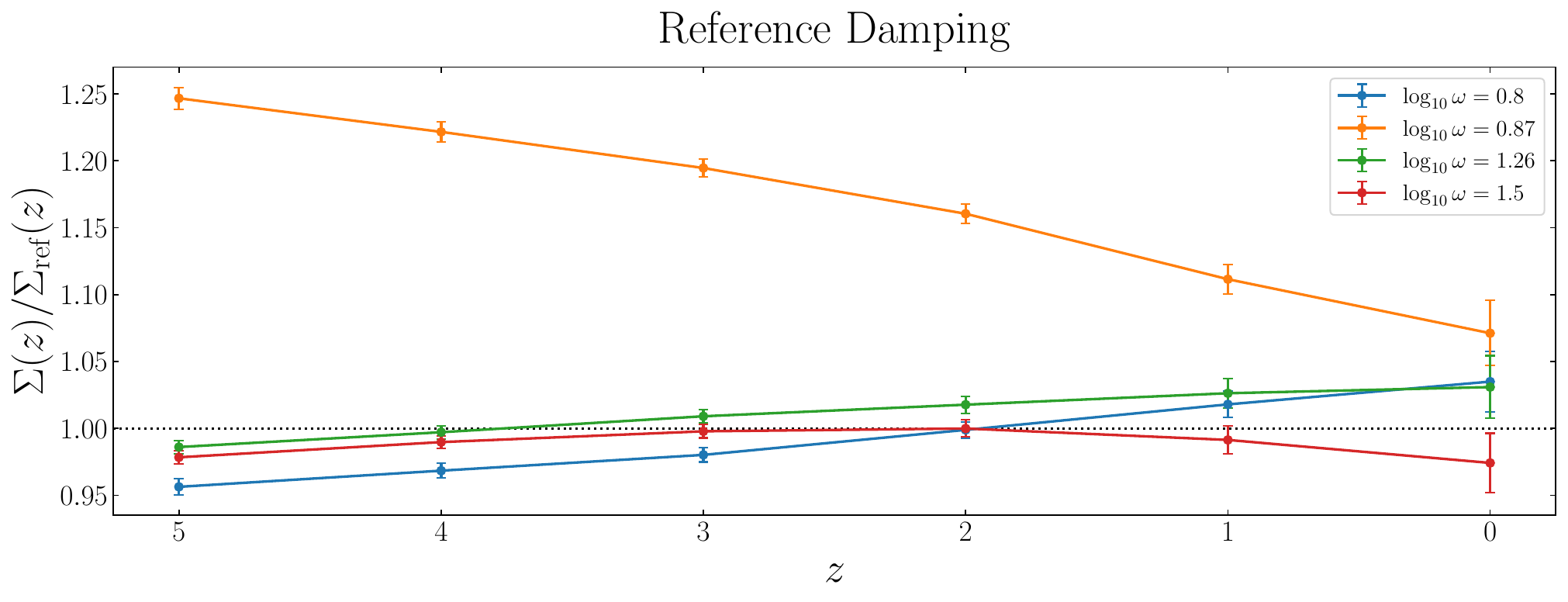}
    \caption{Ratio between the calibrated $\Sigma$ values obtained in this work and those reported by \citet{Ballardini_2020} for logarithmic oscillations.}
    \label{fig:LogOsc_ComparisonRatio}
\end{figure}

As an external consistency check, we compare in Fig.~\ref{fig:LogOsc_ComparisonRatio} our calibrated $\Sigma$ values to those reported by \citet{Ballardini_2020}. The two calibrations agree at the few-percent level across most of the tested domain, with the exception of the $\log_{10} \omega = 0.87$ case.  We note that their $\log_{10}\omega = 0.8$ and $\log_{10}\omega = 0.87$ cases do not follow the otherwise monotonic trend of $\Sigma$ with frequency\footnote{Potentially a mixup between the two frequencies in the presentation of their data?}. While the remaining differences do exceed our statistical uncertainty on $\Sigma(\omega, z)$, they may reflect systematic effects due to differences in simulation setup, fiducial cosmology, or analysis choices. Since~\citet{Ballardini_2020} do not report uncertainties on their best-fitting damping functions, this comparison should be interpreted as a broad consistency check rather than a formal discrepancy test.

\begin{figure}[h]
    \centering
    \includegraphics[width=\linewidth]{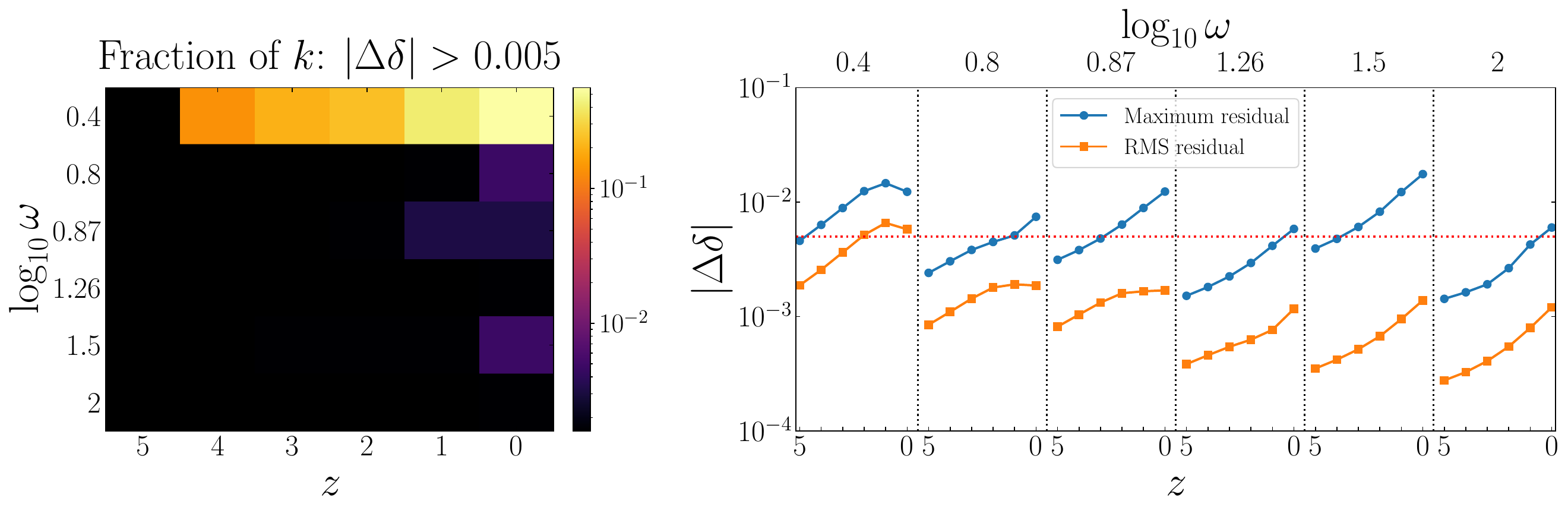}
    \caption{Residual diagnostics for the logarithmic frequency sweep. Left: fraction of Fourier modes whose residual exceeds the adopted threshold. Right: global RMS and maximum residuals as functions of redshift, grouped by oscillation frequency.}
    \label{fig:LogOsc_residual_diagnostics}
\end{figure}

Having calibrated $\Sigma(\omega, z)$, we next test whether the resulting one-parameter damping template reproduces the simulated relative matter power spectra. The residual diagnostics are shown in Fig.~\ref{fig:LogOsc_residual_diagnostics}. For $\log_{10}\omega \geq 0.8$, the fraction of Fourier modes exceeding the threshold is small and the RMS residuals remain controlled below the target accuracy over the full redshift range considered, with only isolated maximum-residual excursions above $5\times10^{-3}$. The lowest frequency case, $\log_{10}\omega = 0.4$, behaves differently. The residuals are small at high redshift, but grow towards late times in the threshold fraction, RMS and maximum residual diagnostics.

Representative fits are shown in Fig.~\ref{fig:LogOsc_trend_plots}, with the full logarithmic residual grid shown in Appendix~\ref{full_results}. For $\log_{10}\omega=2.0$, the calibrated damping template tracks the simulated relative matter power spectrum across the fitted redshift and $k$-range. For $\log_{10}\omega=0.4$, the late-time residuals become visibly larger across the fitted range. The fixed-redshift comparison at $z=0$ shows the same frequency dependence: the higher-frequency cases remain well described by the calibrated template, while the lowest-frequency case develops large residual structure.

\begin{figure}[h]
    \begin{subfigure}[c]{0.99\linewidth}
        \centering
        \includegraphics[width=\linewidth]{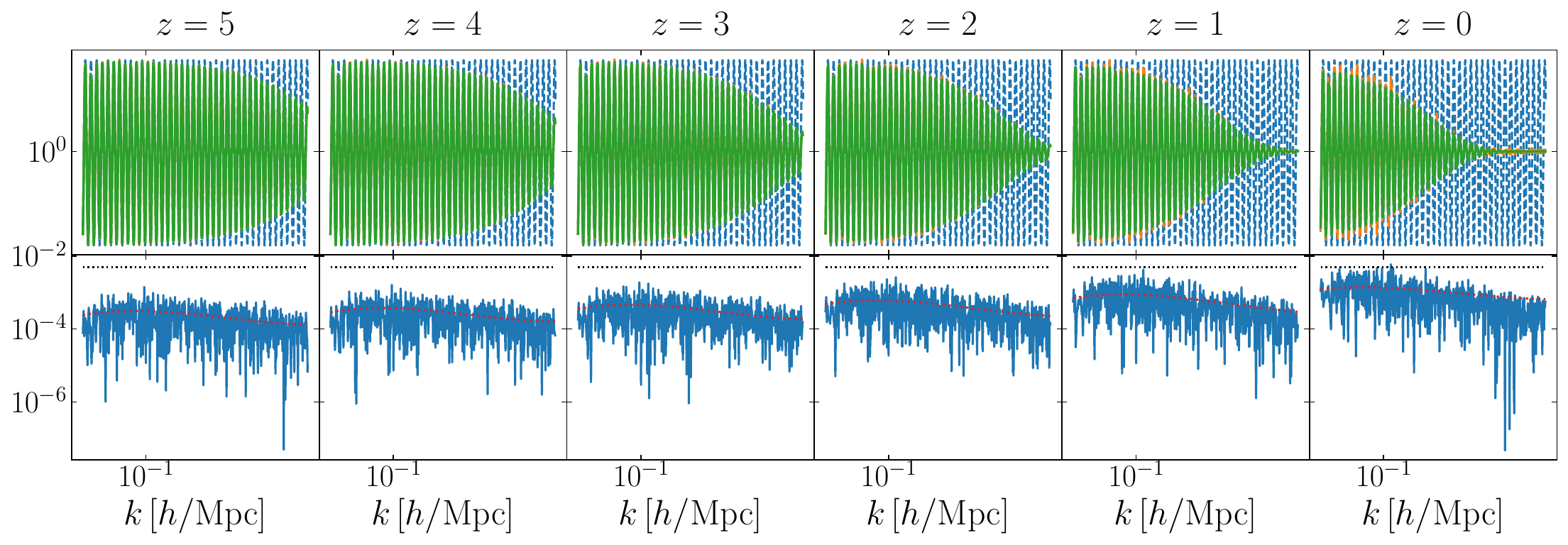}
        \caption{High-frequency logarithmic case, $\log_{10}\omega=2.0$, as a function of redshift.}
        \label{fig:LogOsc_RedshiftTrend2}
    \end{subfigure}
    \hfill
    \begin{subfigure}[c]{0.99\linewidth}
        \centering
        \includegraphics[width=\linewidth]{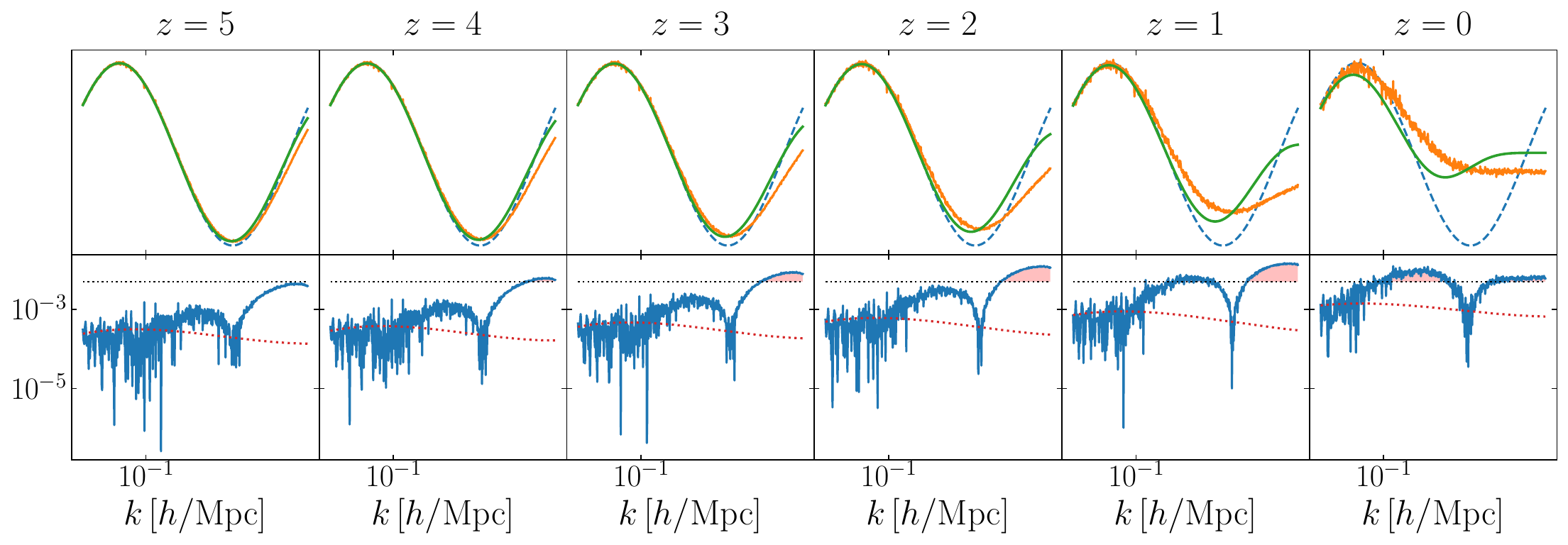}
        \caption{Low-frequency logarithmic case, $\log_{10}\omega=0.4$, as a function of redshift.}
        \label{fig:LogOsc_RedshiftTrend04}
    \end{subfigure}
    \vspace{0.6em}
    \begin{subfigure}[c]{0.99\linewidth}
        \centering 
        \includegraphics[width=\linewidth]{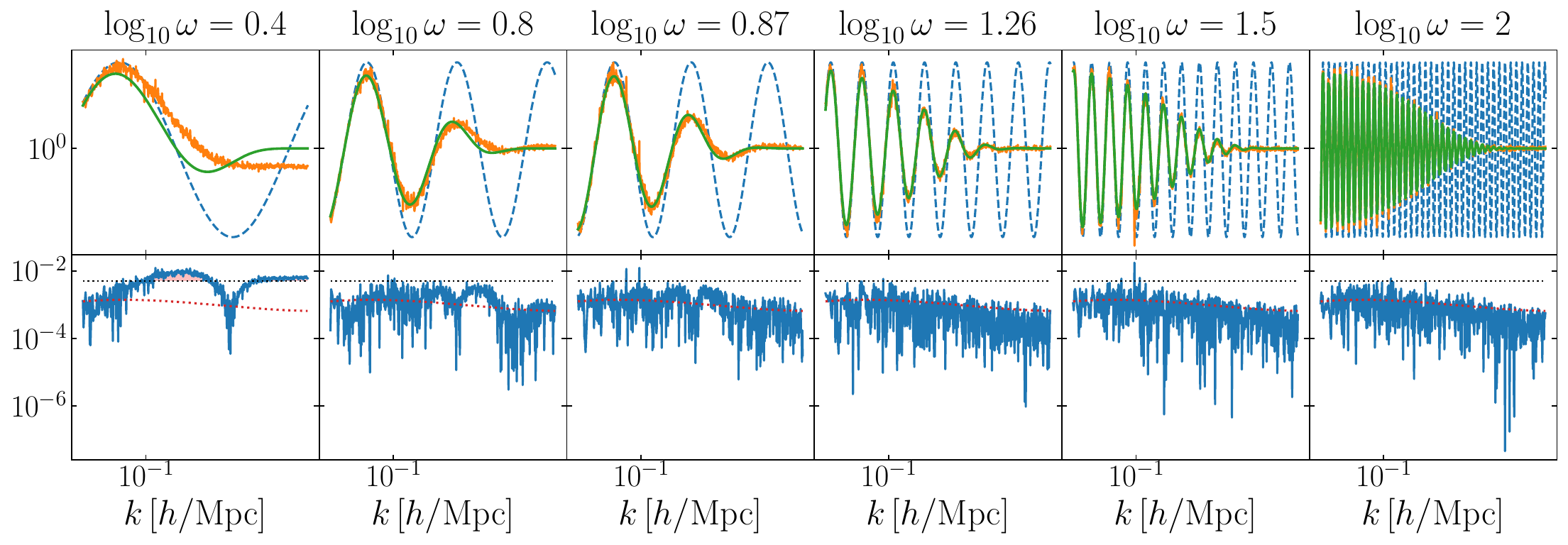}
        \caption{Frequency dependence at fixed redshift, $z=0$}
        \label{fig:LogOsc_FrequencyTrend0}
    \end{subfigure}
    \caption{Representative logarithmic frequency sweep comparisons. In each panel the calibrated semi-analytic template (green) is compared to the simulated relative matter power spectrum (orange) and the linear relative matter power spectrum (blue), with residuals shown underneath.}
    \label{fig:LogOsc_trend_plots}
\end{figure}

\begin{figure}[h]
    \centering
    \includegraphics[width=\linewidth]{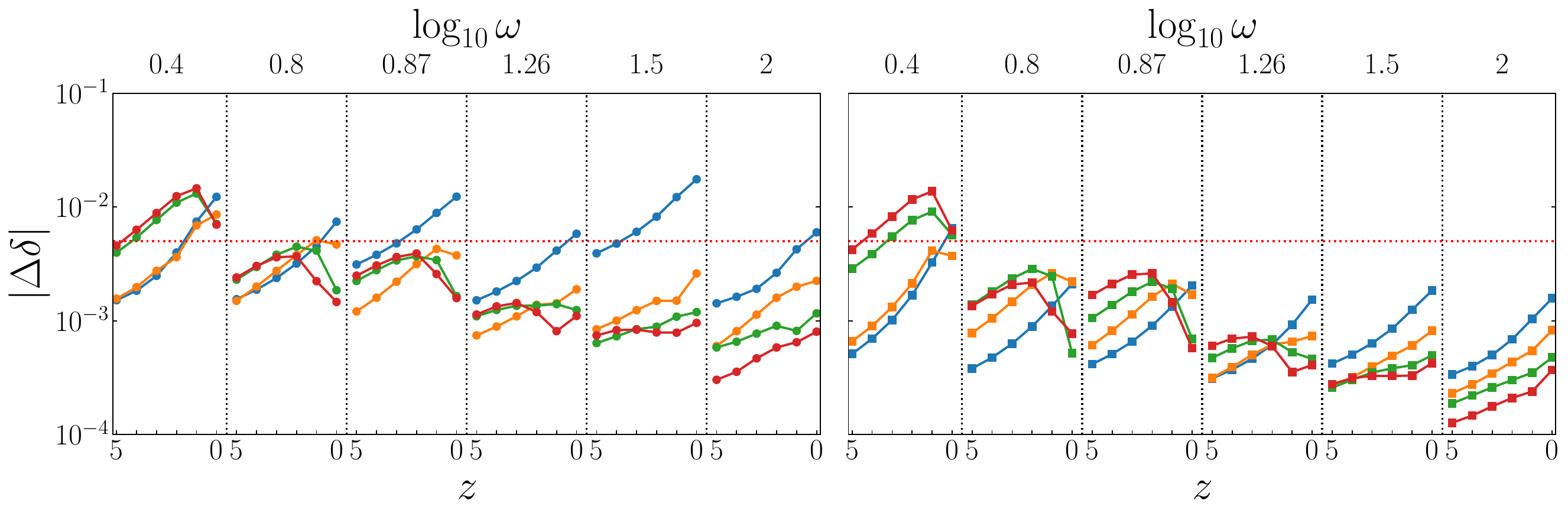}
    \caption{Scale-resolved residual diagnostics for logarithmic oscillations. Left: maximum residual in each $k$-bin, right: RMS residual in each $k$-bin. Colours denote the four $k$-bins used to evaluate the residuals across the fitted range. Blue: $k \in [0.05, 0.19]$, Orange: $k \in [0.19, 0.33]$, Green: $k \in [0.33, 0.46]$, and Red: $k \in [0.46, 0.60]$.}
    \label{fig:LogOsc_binned_residuals}
\end{figure}

The scale-resolved residuals in Fig.~\ref{fig:LogOsc_binned_residuals} show the same pattern. For $\log_{10}\omega \geq 0.8$, the residuals remain small across the fitted wavenumber range, with only mild late-time growth at the smallest scales in isolated deviations. For $\log_{10}\omega=0.4$, both RMS and maximum residuals grow strongly towards low redshift across multiple $k$-bins.  

The logarithmic frequency sweep therefore establishes a clear domain of validity for the semi-analytic damping template. For $\log_{10}\omega \geq 0.8$, the calibrated damping template reproduces the simulated relative matter power spectra to the adopted precision over the full redshift range considered. For $\log_{10}\omega=0.4$, the residuals exceed the adopted accuracy criterion at late times, indicating that the single-parameter Gaussian damping envelope does not provide an appropriate description in this low-frequency regime. In this case, the scale dependence and other structure omitted by the reduced template can no longer be absorbed into one effective value of $\Sigma(\omega, z)$ at the required accuracy. Modelling this case would require a more flexible damping prescription, such as the full leading IR-resummed form or another model with additional degrees of freedom, whose accuracy would itself need to be validated in this regime, and it is therefore excluded from the validated range used in Sec.~\ref{ForecastResults}.

\subsection{Linear Oscillations}
\label{LinResults}
\begin{figure}[h]
    \centering
    \includegraphics[width=\linewidth]{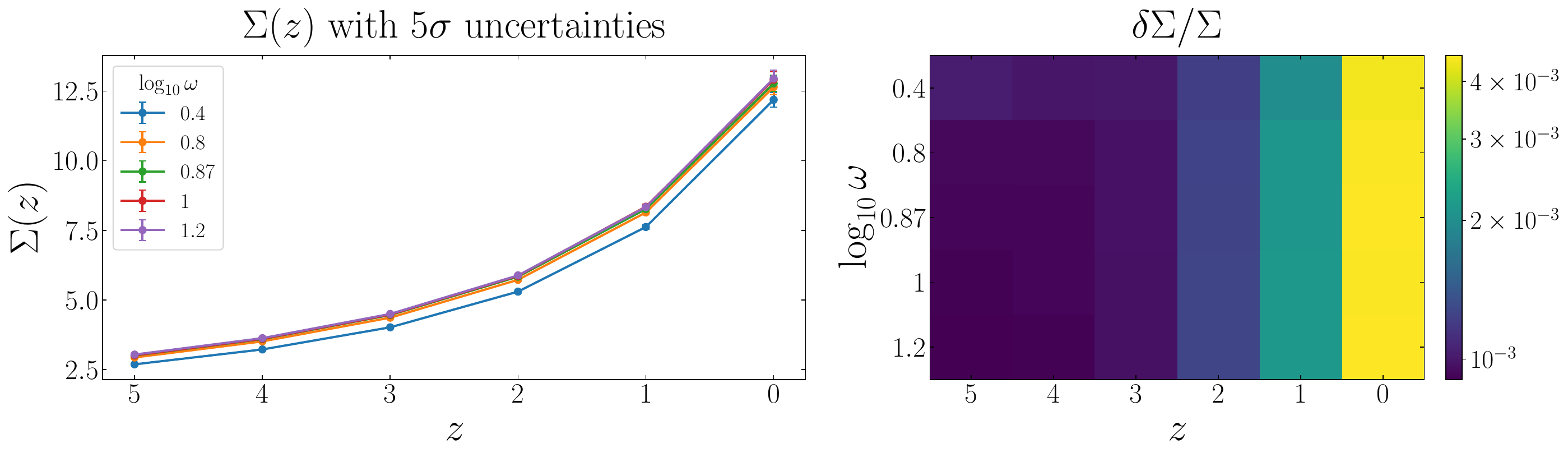}
    \caption{Fit diagnostics for the linear frequency sweep. Left: calibrated $\Sigma(z)$ values. Right: relative uncertainty in the fitted damping function.}
    \label{fig:LinOsc_sigma_uncertianty_pair}
\end{figure}

We next apply the same diagnostics to the linear oscillation model. The calibrated damping functions and their relative uncertainties are shown in Fig.~\ref{fig:LinOsc_sigma_uncertianty_pair}. As in the logarithmic case, $\Delta \Sigma/\Sigma$ remains small across the sampled grid and the fitted $\Sigma$ values increase towards lower redshift. The frequency dependence is weaker than for logarithmic oscillations, with the calibrated $\Sigma(z)$ curves remaining close across the tested frequencies.

\begin{figure}
    \centering
    \includegraphics[width=\linewidth]{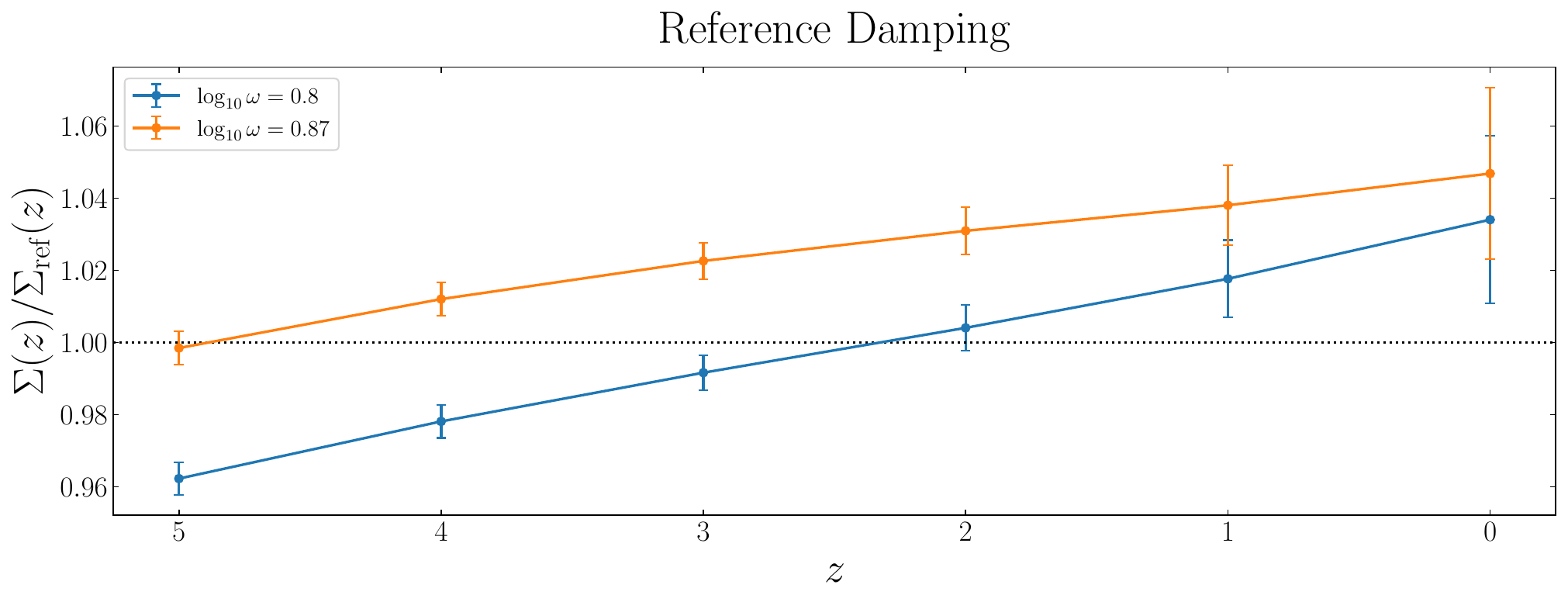}
    \caption{Ratio between the calibrated $\Sigma$ values obtained in this work and those reported by \citet{Ballardini_2020} for linear oscillations with $\phi = 0$.}
    \label{fig:LinOscSigma_Ref}
\end{figure}

Fig.~\ref{fig:LinOscSigma_Ref} compares our calibrated $\Sigma$ values to those reported by \citet{Ballardini_2020} for $\phi = 0$. The two calibrations again agree at the few percent level across the overlapping frequency-redshift range, consistent with the logarithmic comparison.

\begin{figure}[h]
    \centering
    \includegraphics[width=\linewidth]{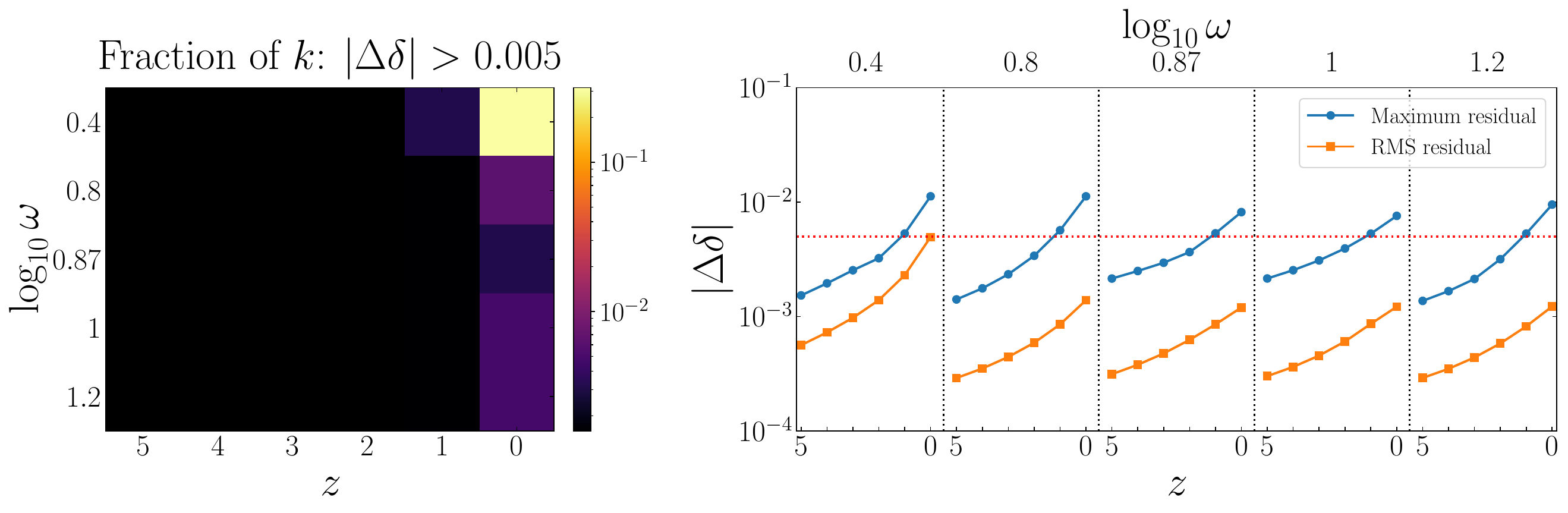}
    \caption{Residual diagnostics for the linear frequency sweep. Left: fraction of Fourier modes whose residual exceeds the adopted threshold. Right: global RMS and maximum residuals as functions of redshift, grouped by oscillation frequency.}
    \label{fig:LinOsc_residual_diagnostics}
\end{figure}

The residual diagnostics are shown in Fig.~\ref{fig:LinOsc_residual_diagnostics}. Although the maximum residual exceeds the threshold at $z=0$, the RMS residual remains below the adopted accuracy threshold across the tested range. Unlike the logarithmic sweep, there is no sharp degradation at the lowest tested frequency.

Representative fits are shown in Fig.~\ref{fig:LinOsc_trend_plots}, with the full linear residual grid shown in Appendix~\ref{full_results}. The residuals grow towards lower redshift, as expected from the increasing importance of non-linear evolution, but remain substantially smaller than the broad late-time residuals found for the lowest-frequency logarithmic case.

\begin{figure} 
    \begin{subfigure}[t]{0.99\linewidth}
        \centering
        \includegraphics[width=\linewidth]{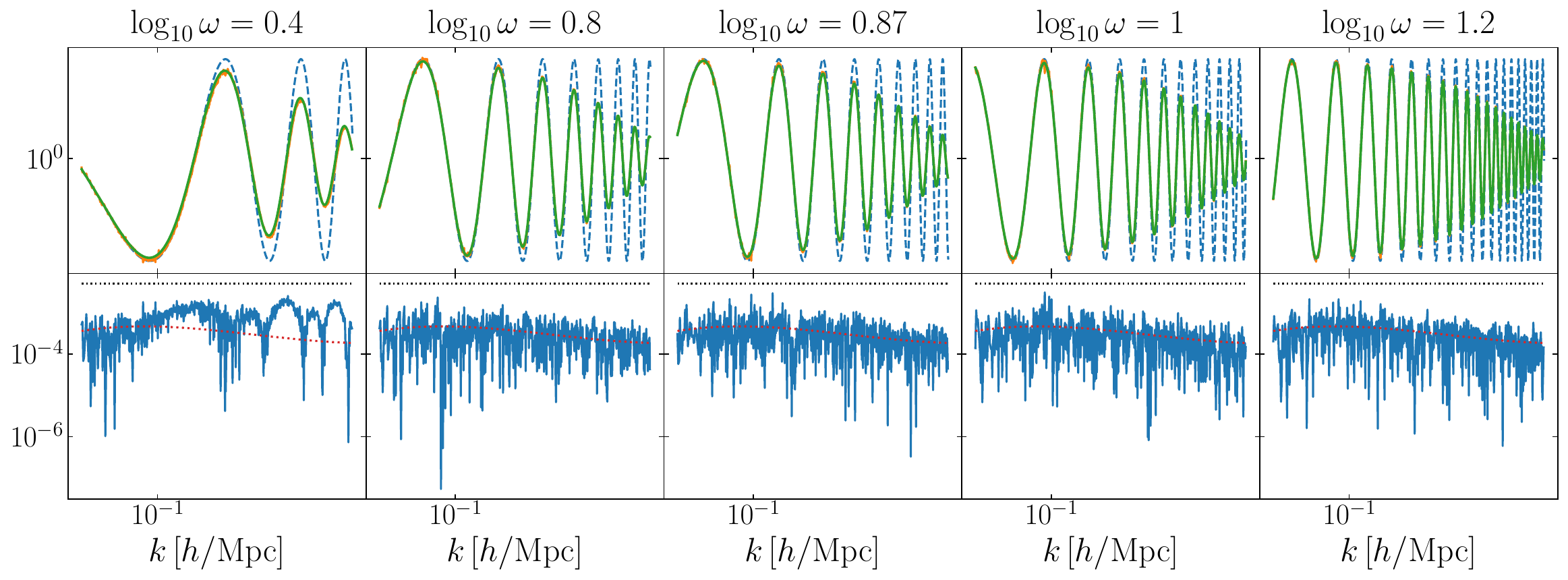}
        \caption{Frequency dependence at fixed redshift, $z=3$.}
        \label{fig:LinOsc_FrequencyTrend}
    \end{subfigure}
    \begin{subfigure}[t]{0.99\linewidth}
        \centering
        \includegraphics[width=\linewidth]{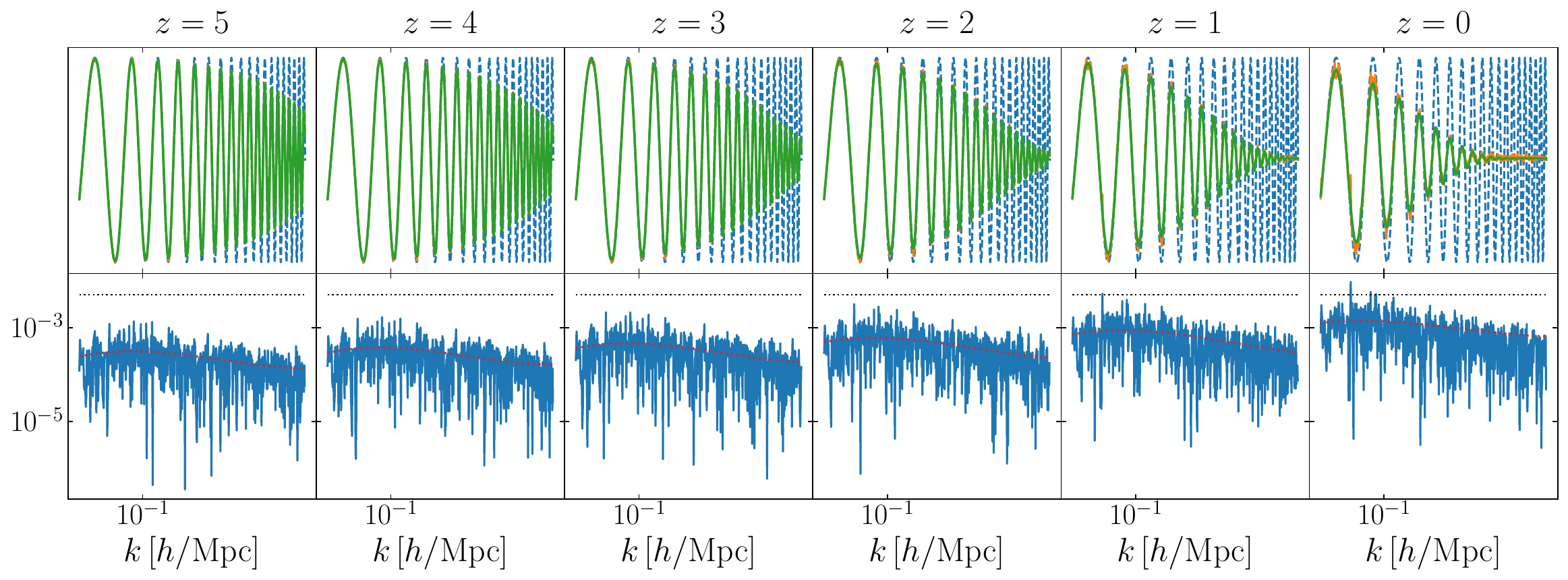}
        \caption{Redshift dependence at fixed frequency, $\log_{10}\omega = 1.2$.}
        \label{fig:LinOsc_RedshiftTrend}
    \end{subfigure}
    \caption{Representative linear frequency sweep comparisons. In each panel the calibrated damping template (green) is compared to the simulated relative matter power spectrum (orange) and linear relative matter power spectrum (blue), with residuals shown underneath.}
    \label{fig:LinOsc_trend_plots}
\end{figure}

\begin{figure}[h]
    \centering
    \includegraphics[width=\linewidth]{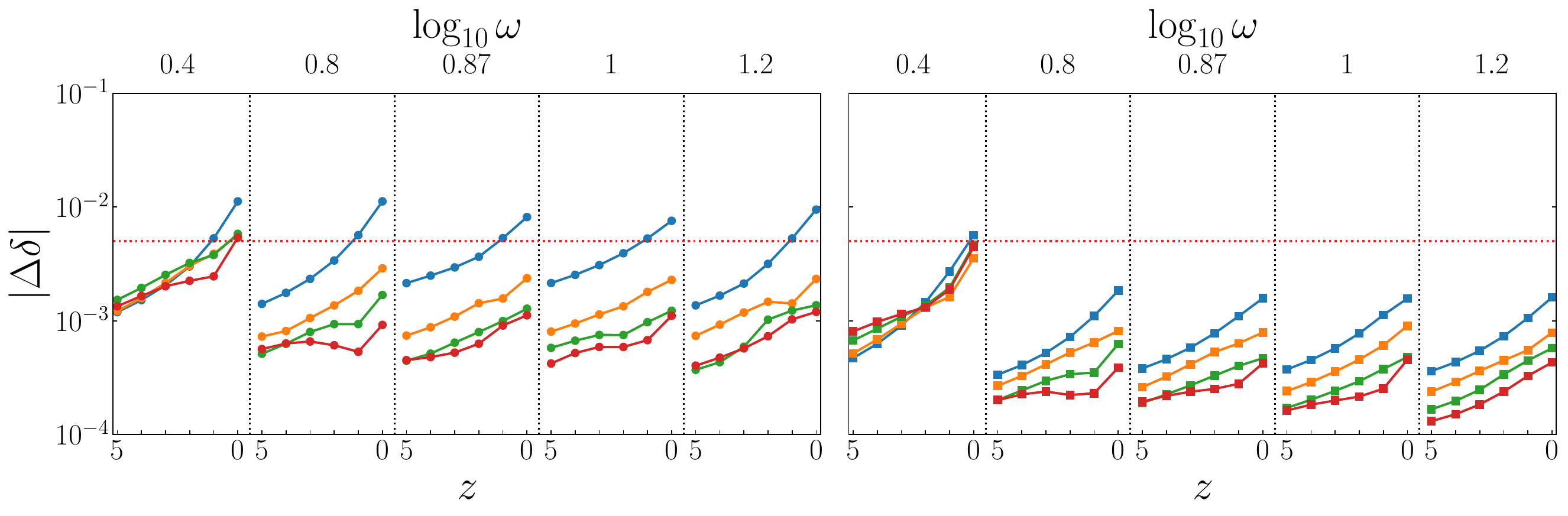}
    \caption{Scale-resolved residual diagnostics for linear oscillations. Left: maximum residuals in each $k$-bin, right: RMS residuals in each $k$-bin. Colours denote the four $k$-bins used to evaluate the residuals across the fitted range. Blue: $k \in [0.05, 0.19]$, Orange: $k \in [0.19, 0.33]$, Green: $k \in [0.33, 0.46]$, and Red: $k \in [0.46, 0.60]$.}
    \label{fig:LinOsc_binned_residuals}
\end{figure}

The scale-resolved residuals in Fig.~\ref{fig:LinOsc_binned_residuals} show that the largest late-time deviations are concentrated in the highest $k$-bins, where non-linear evolution is strongest. The remaining bins stay at or below the adopted accuracy scale in the maximum residual diagnostic, while the RMS residuals remain controlled across the fitted range.

The linear frequency sweep therefore validates the Gaussian damping template across the full tested range,  $0.4 \leq \log_{10}\omega \leq 1.2$. Together with the logarithmic sweep, this extends the validated frequency coverage beyond the specific frequencies calibrated by \citet{Ballardini_2020}, while remaining consistent with their reported $\Sigma$ values where the analyses overlap. These calibrations provide the discrete set of $\Sigma(\omega, z)$ values used to construct the Gaussian-process damping-function emulator.

\subsection{Emulating the Damping Function}
\label{GPRResults}

The simulations calibrate $\Sigma(\omega, z)$ only on a discrete frequency-redshift grid. A likelihood analysis, however, requires the damping template at arbitrary feature frequencies and survey redshifts. We therefore use GPR, as described in Sec.~\ref{GPR}, to turn the calibrated grid into a continuous, uncertainty-aware representation of the damping function. The $\Delta\chi^2=1$ uncertainties are included as training noise, and the calibrated values are tabulated in Appendix~\ref{table_results}.

The damping template depends on the calibrated damping function $\Sigma(\omega, z)$ only through the envelope $\mathcal{D}(k, \Sigma(\omega, z))$, while the remaining feature parameters enter explicitly through the oscillatory part of the template. We therefore emulate $\Sigma(\omega, z)$ rather than the full non-linear matter power spectrum. This keeps the regression low-dimensional, reduces the sparsity of the calibrated grid, and restricts the GP interpolation to the quantity directly constrained by the simulations. It also preserves the modular structure of the model, enabling an existing emulator to supply the smooth non-linear spectrum, while the GP supplies only the additional feature-dependent damping correction.

The emulator does not include an explicit dependence on the primordial phase $\phi$. As discussed in Sec.~\ref{NBodyIC}, \citet{Ballardini_2020} find no significant phase dependence in the calibrated damping function. Likewise, our investigation of the phase dependence in Appendix~\ref{phase_sweep} supports the same conclusion: within the regions of parameter space where the single-parameter damping model is applicable, and at the accuracy relevant for the likelihood analysis below, the calibrated damping function is insensitive to $\phi$. 

We also do not include an explicit dependence on the feature amplitude $\mathcal{A}$ or on the background cosmological parameters. The simulations used for the damping calibration adopt $\mathcal{A} = 0.03$, while the likelihood analysis below varies $\mathcal{A}$ over a narrow range around this value. Changing $\mathcal{A}$ changes the amplitude of the oscillatory feature contribution and therefore acts like a small, scale-dependent perturbation to the matter power spectrum. Over the range of $\mathcal{A}$ values considered here, this produces only a percent-level change in the matter power spectrum, so any change in the non-linear damping of an already percent-level effect encoded by $\Sigma$ is expected to be very small. We therefore do not treat $\mathcal{A}$ as an additional emulator input at the accuracy level considered here. A similar argument applies to the fixed fiducial cosmology. Although the likelihood comparison below does not vary the cosmological parameters, current observations already constrain the relevant parameters at or below the percent level. Variations within the allowed cosmological parameter space volume are therefore expected to produce only small changes in the matter power spectrum over the fitted range, and are not expected to significantly alter the calibrated damping function. 

\begin{figure}
    \centering
    \includegraphics[width=0.99\linewidth]{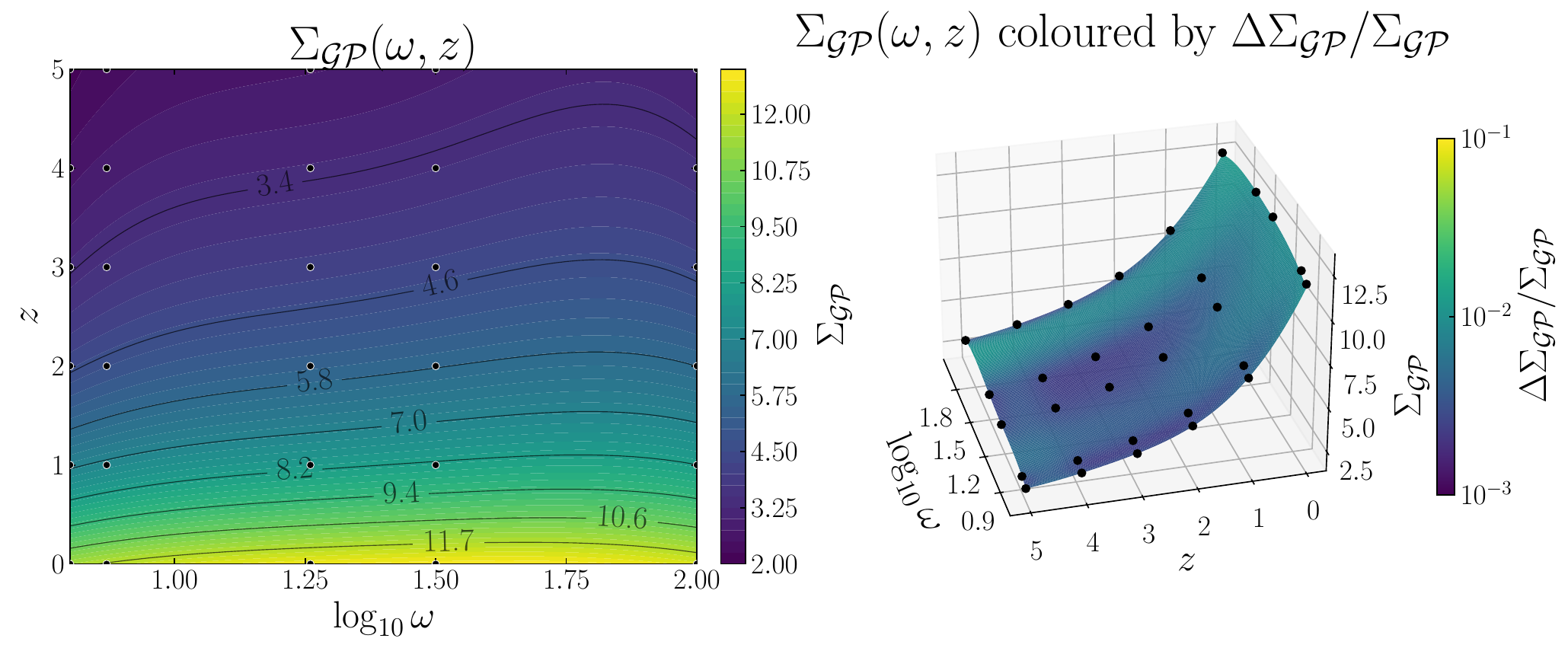}
    \caption{Gaussian-process emulation of the calibrated damping function for logarithmic oscillations. The emulator provides a continuous prediction of $\Sigma(\omega, z)$ and its associated uncertainty across the validated frequency-redshift domain.}
    \label{fig:log_GPRPlots}
\end{figure}

\begin{figure}
    \centering
    \includegraphics[width=0.99\linewidth]{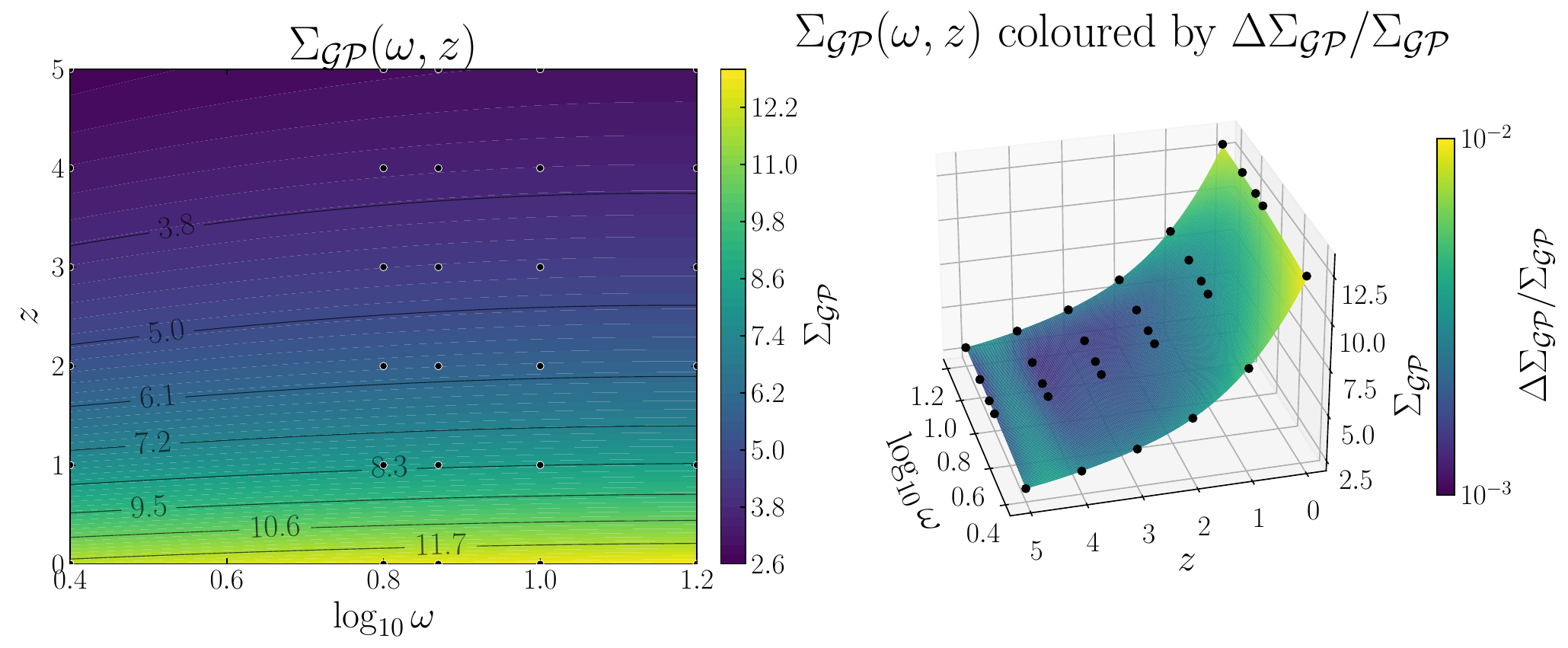}
    \caption{Gaussian-process emulation of the calibrated damping function for linear oscillations. The emulator provides a continuous prediction of $\Sigma(\omega, z)$ and its associated uncertainty across the frequency-redshift domain.}
    \label{fig:Lin_GPRPlots}
\end{figure}

The emulated damping functions are shown in Figs.~\ref{fig:log_GPRPlots} and \ref{fig:Lin_GPRPlots}. For logarithmic oscillations, the emulator is trained only on the validated region, excluding $\log_{10}\omega=0.4$. Within this region, the emulation captures both the increase in $\Sigma$ towards low redshift and the stronger frequency dependence found in the calibration results. For linear oscillations, the emulation is flatter in frequency, reflecting the weaker frequency dependence of the calibrated $\Sigma$ values. The colour scale shows the relative predictive uncertainty, $\Delta\Sigma_{\mathrm{GP}}/ \Sigma_{\mathrm{GP}}$, for visual comparison.

The GP predictive uncertainty remains comparable to the statistical uncertainty of the calibrated training points across the validated region. The simulation grid is therefore adequate for interpolation across the parameter ranges used here, and the GP does not introduce a new dominant source of uncertainty.  In the likelihood analysis below, the absolute predictive uncertainty in $\Sigma$ is propagated through the damping template and included in the diagonal covariance. Its contribution, together with the conservative model-error floor, is subdominant to the survey covariance.

The GPR emulator converts the discrete simulation calibration into a likelihood-ready representation of the damping function. At each likelihood evaluation, the proposed feature frequency and survey redshift determine an interpolated $\Sigma(\omega, z)$, which is then used in $\mathcal{D}(k, \Sigma(\omega, z))$ to suppress the oscillatory component of the matter power spectrum. Since the emulator is already trained, this damping correction adds negligible computational cost compared with the baseline matter-power calculation.

\subsection{Relative Information Comparison with a Euclid-like Spectroscopic Setup}
\label{ForecastResults}

The preceding sections validate the calibrated damping model over a range of feature frequencies and redshifts. We now use this template to quantify how much primordial-feature information can be recovered by modelling the non-linear damping in this way.  This task has two parts: firstly, to estimate how much information is lost due to the damping of the oscillatory signal by non-linear evolution, and secondly, to determine how much of that information can be recovered by using this damping template, rather than discarding the affected scales altogether. To isolate these effects, we use a common Euclid-like spectroscopic clustering covariance and compare three treatments of the same injected logarithmic feature signal.

The primary comparison is between the damped full $k$-range case and the linear only case. In the damped case, we use the full wavenumber range together with the calibrated damping emulator. This retains the modes affected by non-linear evolution, but models the suppression of the oscillatory feature contribution using the validated realistic damping template. In the linear-only case, we keep only modes for which the damping correction is below the adopted accuracy threshold. This corresponds to a conservative analysis in which the affected scales are discarded rather than modelled. We also include a (hypothetical and unphysical) undamped full-range case, in which the damping envelope is set to unity over the same wavenumber range. This provides an idealised upper bound for the information available if the oscillatory feature signal were to survive unchanged.

We restrict this comparison to logarithmic oscillations since for the linear oscillation model, the redshift-dependent linear-only cutoffs leave too few oscillations across the retained wavenumber range to provide a useful comparison of frequency and phase information.

The Euclid-like setup follows the spectroscopic galaxy-clustering configuration used in the primordial features analysis of \citet{EuclidGCsp}, who construct the Fisher information from the observed three-dimensional galaxy power spectrum, $P_{\mathrm{obs}}(k, \mu, z)$, summed over wavenumber, angle to the line of sight, and redshift bins. Here we use a simplified isotropic version of this structure. We retain Euclid-like redshift bins, number densities, galaxy biases, survey volumes, and finite $k$-binning, but omit redshift-space distortions, Alcock-Paczynski factors, redshift errors, Finger-of-God damping, and the full survey window. The result should therefore be interpreted as a relative information comparison between the three treatments, not as a realistic forecast of absolute sensitivities.

We use four redshift bins centred at
\begin{align}
    z=\{1.0, 1.2, 1.4, 1.65\},
\end{align}
with widths $\Delta z=0.2$ for the first three bins and $\Delta z=0.3$ for the highest-redshift bin, following the Euclid GCsp primordial-features setup. The corresponding H$\alpha$ galaxy number densities and linear biases are those used in the same setup. The full-range likelihood uses
\begin{align}
    0.05  h \, \mathrm{Mpc}^{-1} \leq k \leq 0.25 \: h \, \mathrm{Mpc}^{-1},
\end{align}
matching the pessimistic upper bound used in \cite{EuclidGCsp} and the lower bound used in the simulation validation. The model is evaluated on a fine internal $k$ grid and then averaged into top-hat bandpowers with
\begin{align}
    \Delta k  = 0.004 \: h\,\mathrm{Mpc}^{-1}.
\end{align}
This avoids treating isolated theory evaluations as the measured quantities. A full survey-window convolution would mix neighbouring bandpowers and require a non-diagonal covariance; the top-hat average is therefore used as a diagonal-covariance analogue of finite $k$-resolution.

For each redshift bin, the featureless linear matter power spectrum, $\mathcal{P}^{0}_{\mathrm{m, lin}}$, is computed once and cached. At each likelihood evaluation, the proposed feature parameters determine the linear relative feature contribution. For logarithmic oscillations this is
\begin{align}
    \delta_{\mathrm{lin}}(k, \mathcal{A}, \omega, \phi) = \mathcal{A}\cos\left[\omega \log \frac{k}{k_{\ast}} + \phi \right],
\end{align}
where $\mathcal{A}$ is the feature amplitude, $\omega$ is the logarithmic feature frequency, and $\phi$ is the phase.

For the damped case, the GPR emulator supplies $\Sigma_{\mathrm{GP}}(\omega, z)$, giving,
\begin{align}
    \delta_{\mathrm{pred}}(k, z) = \delta_{\mathrm{lin}}(k, \mathcal{A}, \omega, \phi) \exp\left[-\frac{k^2 \Sigma^2_{\mathrm{GP}}(\omega, z)}{2}\right].
    \label{eq:forecsat_delta_pred}
\end{align}
The corresponding predicted matter power spectrum is
\begin{align}
    \mathcal{P}_{\mathrm{m, pred}}(k, z) = \mathcal{P}^{0}_{\mathrm{m, lin}}(k, z)\left[1 + \delta_{\mathrm{pred}}(k, z) \right].
    \label{eq:forecast_power_pred}
\end{align}
For the undamped full-range case, the same expression is used with the exponential damping envelope set to unity. In a full likelihood comparison, $\mathcal{P}^{0}_{\mathrm{m, lin}}$ would be replaced by a non-linear prediction for the smooth reference spectrum. This is not expected to affect the relative comparison performed here, since the smooth non-linear correction is common to the three cases and will not be correlated with any of the features parameters.  The diagnostic is instead controlled by how the feature contribution is damped.

For the linear-only case, the damped model is retained, but the likelihood is restricted to wavenumbers for which the maximum damping-induced change in the relative matter power spectrum remains below the residual validation threshold. Specifically, for each redshift bin we define $k_{\max}^{\mathrm{l}}(z)$ by requiring
\begin{align}
    \mathcal{A}_{\mathrm{fid}}\left[1 - \exp\left(-\frac{k^2 \Sigma_{\max}^2(z)}{2} \right) \right] < 5\times10^{-3},
\end{align}
where $\Sigma_{\max}(z)$ is the maximum emulator-predicted damping function over the validated frequency range at each redshift. This gives
\begin{align}
    k_{\max}^{\mathrm{l}}(z) = \{0.106, 0.115, 0.123, 0.134\} \: h\,\mathrm{Mpc}^{-1}
\end{align}
for $z=\{1.0, 1.2, 1.4, 1.65\}$, respectively, for the trained GP used in these tests. The linear-only comparison therefore keeps only modes for which the calibrated non-linear correction is smaller than the same residual tolerance used to validate the damping template.

The data vector is generated from the damped model at the fiducial parameters
\begin{align}
    \mathcal{A}_{\mathrm{fid}} = 0.03,
    \qquad 
    \log_{10}\omega_{\mathrm{fid}} =1.26,
    \qquad 
    \phi_{\mathrm{fid}} = \pi.
\end{align}
As discussed in Sec.~\ref{GPRResults}, the emulator is trained at fixed $\mathcal{A} = 0.03$ and $\phi=0$, but no significant dependence of the damping function on either parameter is expected at the accuracy used here.

The likelihood covariance is diagonal and contains three contributions. The first is the Euclid-like Gaussian bandpower uncertainty, including survey volume, shot noise and sample variance. The second is the uncertainty propagated from the GP prediction for $\Sigma_{\mathrm{GP}}$. From Eq.~\ref{eq:forecast_power_pred}, the corresponding power-spectrum uncertainty is approximated by
\begin{align}
    \sigma^2_{P, \mathrm{GP}} 
    = 
    \left[
        \mathcal{P}^{0}_{\mathrm{m, lin}}(k, z)
        \delta_{\mathrm{pred}}(k, z) 
        k^2 
        \Sigma_{\mathrm{GP}}(\omega, z) 
        \sigma_{\Sigma, \mathrm{GP}}(\omega, z)
        \right]^2 .
\end{align}
The third contribution is a conservative model-error floor corresponding to the residual validation threshold, $5\times10^{-3}$, applied in relative-spectrum units and converted to matter-power units. This prevents the likelihood from assuming that the damping template is more accurate than demonstrated by the simulation validation. For this setup, both the GP uncertainty and the model-error floor are subdominant to the survey covariance.

\begin{figure}[h]
    \centering
    \includegraphics[width=0.8\linewidth]{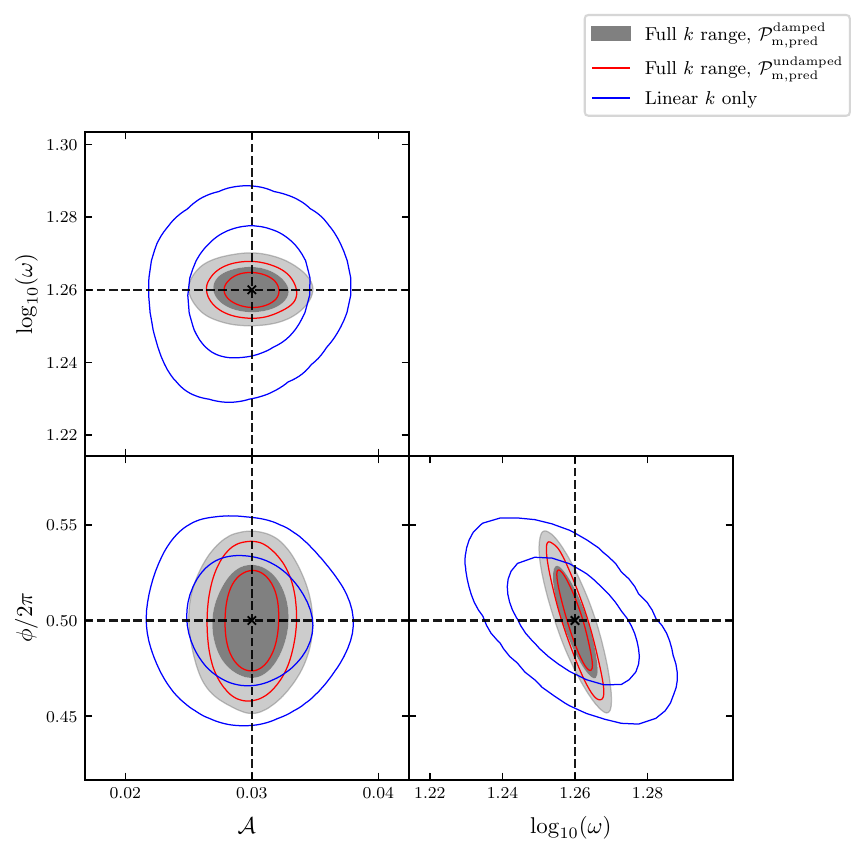}
    \caption{Relative-information comparison for the injected logarithmic feature signal. The damped case uses the calibrated Gaussian damping emulator over the full $k$-range, the undamped case sets the damping envelope to unity over the same range, and the linear-only case restricts the likelihood to modes for which the damping correction is below the adopted residual threshold.}
    \label{fig:wiggle_comparison_plot}
\end{figure}

The posterior constraints are shown in Fig.~\ref{fig:wiggle_comparison_plot}, with marginal constraints summarised in Table~\ref{tab:wiggle_comparison_constraints}. The absolute numerical values should not be interpreted as a Euclid forecast, since the likelihood fixes the background cosmology and uses a simplified isotropic covariance. The relative behaviour is the intended diagnostic. The undamped full-range gives the tightest constraints because it assumes that the oscillatory information survives unchanged across the full fitted range. The damped full-range case gives weaker constraints because non-linear evolution suppresses part of the feature signal. The linear-only case is weaker again because it discards the scales where non-linear modelling would be required. 

\begin{table}[h]
\centering
\begin{tabular}{lccc}
\hline
Scenario & $\mathcal{A}$ & $\log_{10}\omega$ & $\phi/2\pi$ \\
\hline
Full $k$ range, $\mathcal{P}_{\mathrm{m,pred}}^{\mathrm{damped}}$ & $0.0299^{+0.0020}_{-0.0019}$ & $1.2600^{+0.0040}_{-0.0040}$ & $0.4993^{+0.0190}_{-0.0189}$ \\
Full $k$ range, $\mathcal{P}_{\mathrm{m,pred}}^{\mathrm{undamped}}$ & $0.0300^{+0.0014}_{-0.0014}$ & $1.2600^{+0.0031}_{-0.0033}$ & $0.4999^{+0.0170}_{-0.0172}$ \\
Linear $k$ only & $0.0297^{+0.0032}_{-0.0032}$ & $1.2597^{+0.0112}_{-0.0117}$ & $0.4995^{+0.0219}_{-0.0200}$ \\
\hline
\end{tabular}
\caption{Marginal constraints from the three relative-information comparison scenarios.}
\label{tab:wiggle_comparison_constraints}
\end{table}

The comparison quantifies the information recovered by modelling the non-linear damping. For the amplitude, the damped full-range case gives $\sigma(\mathcal{A}) \simeq 0.0020$, compared with  $0.0032$ in the linear-only case and $0.0014$ in the ideal undamped case. The calibrated damping template therefore recovers a substantial fraction of the amplitude information lost under the conservative linear-only cutoff, while remaining within a factor of order unity of the ideal full-range case. For the frequency, the gain is larger, the calibrated damping model improves $\sigma(\log_{10}\omega)$ by nearly a factor of three relative to the linear-only case, while remaining close to the ideal undamped result. The phase constraint changes more mildly, but follows the same ordering.

This simplified likelihood is, as expected, more constraining than the pessimistic GCsp-only Fisher result of \citep{EuclidGCsp}, which reports marginal uncertainties of approximately $25\%$ on $\mathcal{A}$, $4.7\%$ on $\omega$ and $0.053$ on the normalised phase. This is likely due to our simplified setup, with the background cosmology kept fixed and the omission of the full anisotropic modelling. 
 
\section{Conclusion}
\label{conclusion}

We have presented a quantitative validation of the compact model proposed by \citet{Ballardini_2020} for describing the non-linear suppression of primordial wiggles in the late-time matter power spectrum. The model retains the leading Gaussian damping structure obtained from IR-resummed perturbation theory~\citep{Blas2016, Beutler2019, Vasudevan2019}, but compresses the relevant non-linear dynamics into a single effective damping scale, $\Sigma(\omega, z)$, fitted to $N$-body simulations. Over the range $0.05\; h \mathrm{Mpc}^{-1} \leq k \leq 0.6 \; h \mathrm{Mpc}^{-1}$, we find that this Gaussian damping envelope 
\begin{align*}
    \mathcal{D}(k, \Sigma(\omega, z)) = e^{-k^2 \Sigma^2(\omega, z)/2}
\end{align*}
provides an accurate description of the simulated non-linear suppression across a broad region of the logarithmic and linear oscillatory feature space.

Using matched feature and featureless $N$-body simulation pairs, together with an independent estimate of the simulation variance, we calibrated the damping function $\Sigma(\omega, z)$ and quantified both its statistical uncertainty and the residual error of the damping template. For logarithmic oscillations with $0.8 \leq \log_{10}\omega \leq 2.0 $, and for all linear oscillations considered, $0.4 \leq \log_{10}\omega \leq 1.2$, the damping function is well determined and the residuals remain controlled at or below our accuracy target of $5\times10^{-3}$. Towards lower logarithmic oscillation frequencies, the one-parameter damping model becomes less accurate, with the residual error exceeding our residual threshold at larger $k$. This behaviour reflects the limitations of compressing the more general scale-dependent logarithmic-feature evolution into one $k$-independent damping scale. Within the validated frequency ranges, however, the leading displacement-driven suppression can be represented by this compact template without exceeding the accuracy target adopted for LSS analyses.

Caution should therefore be exercised for $\log_{10}\omega < 0.8$ in the logarithmic template. However, at larger frequencies the damping model is sufficiently accurate for feature analyses targeting the percent-level sensitivity of large-scale structure surveys.

We then used Gaussian Process Regression to turn the discrete simulation calibration into a continuous emulator for $\Sigma(\omega, z)$. This emulator preserves the structure of the semi-analytic template while allowing the damping function and its predictive uncertainty to be evaluated inside a likelihood. As a proof of concept, we inserted the emulator into a fixed-cosmology Euclid-like GCsp information comparison for logarithmic wiggles, including survey sensitivity, GP predictive uncertainty, and a conservative model-error floor tied to the $5 \times 10^{-3}$ residual threshold. The resulting comparison shows that modelling the damping recovers a substantial fraction of the constraining power that would otherwise be discarded by a conservative linear-only cut. For the feature amplitude, the damped full-range case reduces the marginal uncertainty by approximately $40\%$ relative to the linear-only case. For the frequency, the improvement is larger, with the marginal uncertainty in $\log_{10}\omega$ reduced by approximately $65\%$. The phase constraint changes more mildly, improving by approximately $10\%$.

The validation in this work was performed at fixed fiducial cosmology, and the likelihood comparison likewise fixes the background cosmological parameters. Since currently allowed variations in these parameters are small, the induced changes in $\Sigma$ are expected to be subdominant for the comparison considered here, although this dependence has not been explicitly calibrated. A full survey analysis should nevertheless either validate this cosmology dependence directly, include it as an additional modelling uncertainty, or extend the emulator to include cosmological parameters.

These results establish a practical simulation-calibrated route from primordial wiggles models to non-linear large scale structure likelihoods. A natural extension would be to replace the fixed simulation grid with an adaptive design strategy. In higher-dimensional spaces, for example if amplitude or cosmological parameters are included, a grid-based simulation campaign scales poorly with dimension. Bayesian optimisation, such as \texttt{BOBE}~\citep{Cohen_2026}, or related active-learning methods could instead select new $N$-body simulations according to where they are expected to most improve the damping emulator, reducing the number of simulations required to reach a target accuracy. 

\section*{Acknowledgements}
NC acknowledges support from
an Australian Government Research Training Program Scholarship. This research includes computations using the computational cluster Katana supported by Research Technology Services at UNSW Sydney

\section*{Code Availability}
The code used for the calibration, Gaussian Process Regression, plotting, and forecast analysis is available at \url{https://github.com/nathyglobal/WigglesGP}.

\bibliographystyle{apsrev}
\bibliography{bib}

\begin{appendix}
\section{Phase dependence of the damping calibration}
\label{phase_sweep}
The analysis in Sec.~\ref{ForecastResults} varies both the oscillation frequency and phase, while the damping emulator is trained as a function of frequency and redshift only, $\Sigma(\omega, z)$. This assumes that, within the validated domain of the semi-analytic template, the calibrated damping function is effectively independent of the primordial phase. In this appendix we test that assumption by repeating the calibration for several values of $\phi$.

We first consider phase sweeps within the validated region. For logarithmic oscillations, we use $\log_{10}\omega=1.26$, which lies within the domain used in the likelihood comparison. For linear oscillations, we use $\log_{10}\omega=0.4$, the lowest-frequency validated linear template and therefore the case expected to be most sensitive to phase shifts. In each case we use $\phi/\pi = \{0, 0.5, 0.75, 1.0 \}$ and apply the same calibration procedure used in the main analysis.

\begin{figure}[h]
    \centering
    \includegraphics[width=\linewidth]{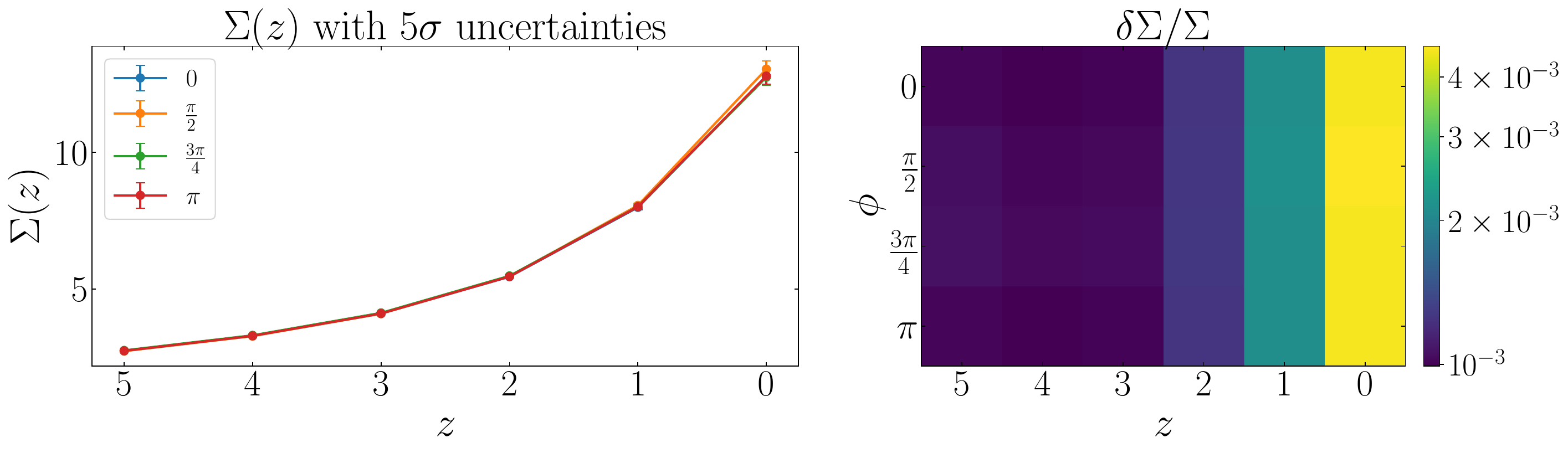}
    \caption{Phase dependence diagnostics for the logarithmic phase sweep at $\log_{10}\omega=1.26$. Left: calibrated $\Sigma(z)$ values for each phase. Right: relative uncertainty in the fitted damping function.}
    \label{fig:LogOsc_phase_sigma_uncertainty_w126}
\end{figure}

For the logarithmic case, shown in Fig.~\ref{fig:LogOsc_phase_sigma_uncertainty_w126}, the calibrated $\Sigma(z)$ curves are stable under phase variation, and the relative fit uncertainties remain small. This is consistent with the results of \citet{Ballardini_2020} and further justifies the conclusion that the calibrated damping function is insensitive to $\phi$ for this validated logarithmic template.

\begin{figure}[h]
    \centering
    \includegraphics[width=\linewidth]{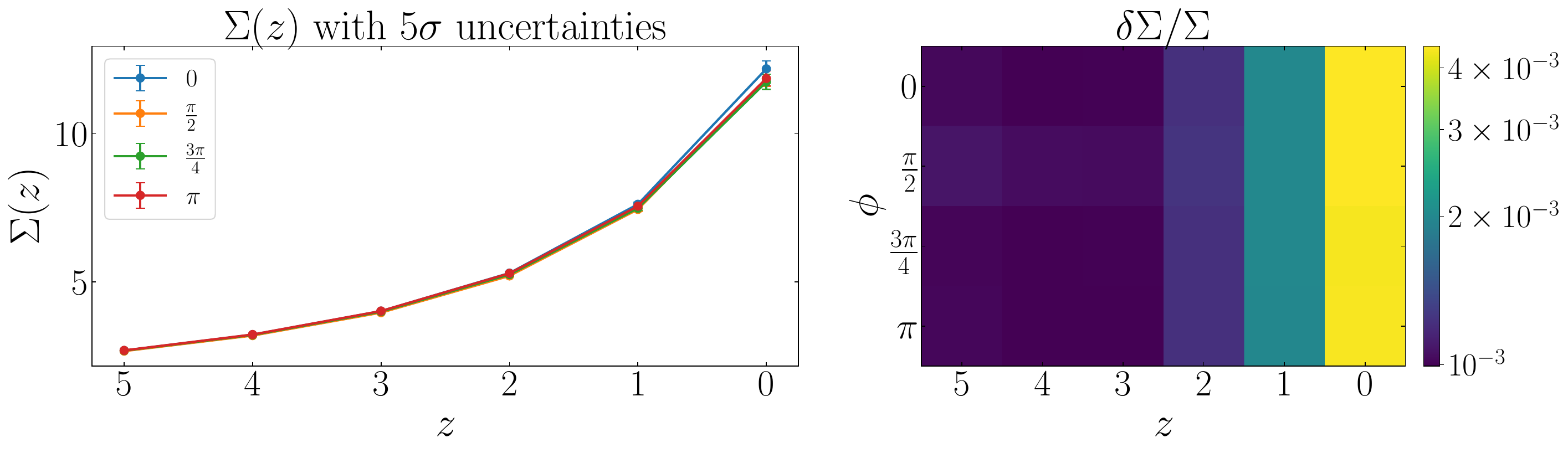}
    \caption{Phase dependence diagnostics for the linear phase sweep at $\log_{10}\omega=0.4$. Left: calibrated $\Sigma(z)$ values for each phase. Right: relative uncertainty in the fitted damping function.}
    \label{fig:LinOsc_phase_sigma_uncertainty_w04}
\end{figure}

For the linear case, shown in Fig.~\ref{fig:LinOsc_phase_sigma_uncertainty_w04}, the calibrated $\Sigma(z)$ curves are also effectively unchanged as the phase is varied and the fit uncertainties remain uniformly small. Since this is the lowest-frequency validated linear template, it provides the most conservative phase-sensitivity test for the linear frequency sweep.

\begin{figure}[h]
    \centering
    \includegraphics[width=\linewidth]{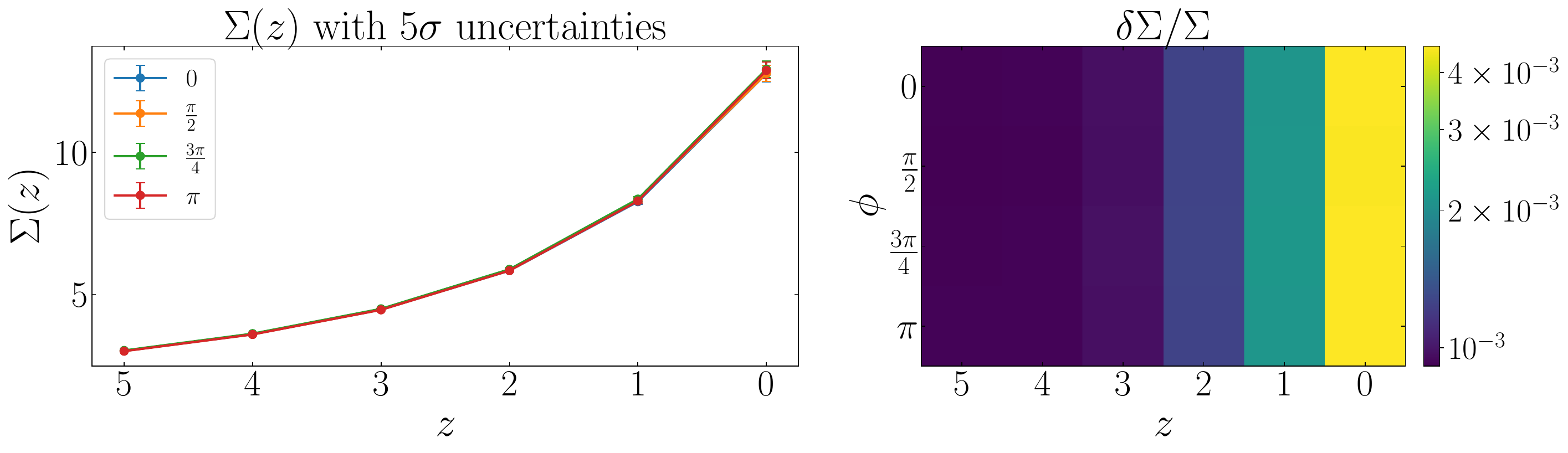}
    \caption{Phase dependence diagnostics for the linear phase sweep at $\log_{10}\omega=0.87$. Left: calibrated $\Sigma(z)$ values for each phase. Right: relative uncertainty in the fitted damping function.}
    \label{fig:LinOsc_phase_sigma_uncertainty_w087}
\end{figure}

As an additional check, we repeat the linear phase sweep at $\log_{10}\omega=0.87$, shown in Fig.~\ref{fig:LinOsc_phase_sigma_uncertainty_w087}. The same behaviour is found, varying $\phi$ does not produce a measurable shift in the calibrated damping function.

\begin{figure}[h]
    \centering
    \includegraphics[width=\linewidth]{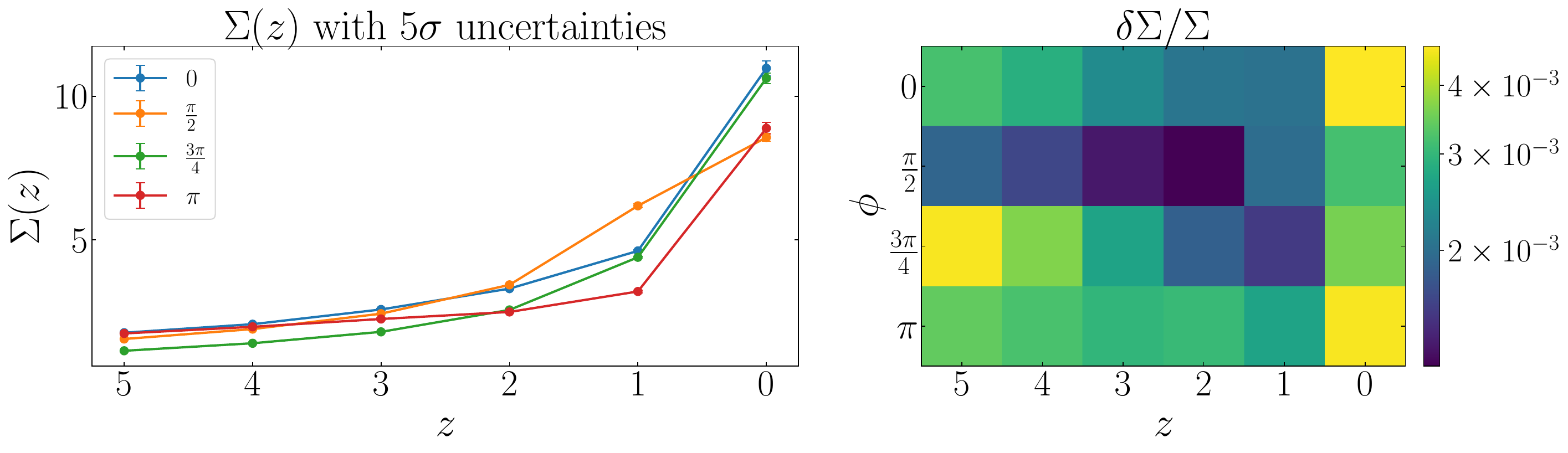}
    \caption{Phase dependence diagnostics for the logarithmic phase sweep at $\log_{10}\omega=0.4$. Left: calibrated $\Sigma(z)$ values for each phase. Right: relative uncertainty in the fitted damping function. This frequency lies outside the validated domain of the single-envelope template.}
    \label{fig:LogOsc_phase_sigma_uncertainty_w04}
\end{figure}

We also test the low-frequency logarithmic case, $\log_{10}\omega=0.4$, shown in Fig.~\ref{fig:LogOsc_phase_sigma_uncertainty_w04}. The recovered $\Sigma(z)$ values show a visible dependence on phase, but this occurs in the regime where the single-envelope damping template does not accurately reproduce the simulated spectra. It should therefore not be interpreted as evidence that the damping emulator must generally include $\phi$ as an input. Rather, once the one-parameter damping template no longer provides a good description of the non-linear distortion, the best-fitting value of $\Sigma$ can become sensitive to how the broad oscillatory feature is positioned within the fitted $k$-range.

These phase sweeps support the use of a damping emulator trained on $\Sigma(\omega, z)$ rather than $\Sigma(\omega, \phi, z)$,  within the validated domain of the semi-analytic model. The calibrated damping function shows no significant phase dependence for the validated logarithmic and linear tests, including the lowest-frequency validated linear case where phase sensitivity is expected to be largest. The only visible phase dependence occurs for the low-frequency logarithmic template, which is already outside the regime where the single-envelope damping template is valid. 

\newpage
\section{$\Sigma(z)$ Fitted to $N$-body relative power spectra}
\label{table_results}
\begin{table}[h]
\centering
\begin{tabular}{ccccccc}
\hline
$\log_{10}\omega$ & $z=5$ & $z=4$ & $z=3$ & $z=2$ & $z=1$ & $z=0$ \\
\hline
$0.4$ & $1.767 \pm 0.006$ & $2.063 \pm 0.006$ & $2.575 \pm 0.006$ & $3.301 \pm 0.007$ & $4.615 \pm 0.009$ & $10.973 \pm 0.052$ \\
$0.8$ & $2.171 \pm 0.003$ & $2.624 \pm 0.003$ & $3.303 \pm 0.004$ & $4.465 \pm 0.005$ & $6.800 \pm 0.013$ & $11.695 \pm 0.051$ \\
$0.87$ & $2.381 \pm 0.003$ & $2.859 \pm 0.003$ & $3.572 \pm 0.004$ & $4.746 \pm 0.006$ & $7.058 \pm 0.014$ & $12.029 \pm 0.055$ \\
$1.26$ & $2.731 \pm 0.003$ & $3.281 \pm 0.003$ & $4.097 \pm 0.004$ & $5.445 \pm 0.007$ & $7.984 \pm 0.017$ & $12.783 \pm 0.058$ \\
$1.5$ & $2.916 \pm 0.003$ & $3.494 \pm 0.003$ & $4.350 \pm 0.004$ & $5.739 \pm 0.007$ & $8.248 \pm 0.018$ & $12.859 \pm 0.058$ \\
$2.0$ & $2.984 \pm 0.003$ & $3.559 \pm 0.003$ & $4.408 \pm 0.004$ & $5.759 \pm 0.007$ & $8.151 \pm 0.017$ & $12.446 \pm 0.055$ \\
\hline
\end{tabular}
\caption{$\Sigma(z)$ with uncertainties for logarithmic oscillations.}
\label{tab:LogOsc_sigma_z_grid}
\end{table}

\begin{table}[h]
\centering
\begin{tabular}{ccccccc}
\hline
$\log_{10}\omega$ & $z=5$ & $z=4$ & $z=3$ & $z=2$ & $z=1$ & $z=0$ \\
\hline
$0.4$ & $2.688 \pm 0.003$ & $3.223 \pm 0.003$ & $4.017 \pm 0.004$ & $5.299 \pm 0.006$ & $7.620 \pm 0.015$ & $12.187 \pm 0.054$ \\
$0.8$ & $2.935 \pm 0.003$ & $3.511 \pm 0.003$ & $4.363 \pm 0.004$ & $5.723 \pm 0.007$ & $8.141 \pm 0.017$ & $12.646 \pm 0.057$ \\
$0.87$ & $3.005 \pm 0.003$ & $3.593 \pm 0.003$ & $4.459 \pm 0.004$ & $5.835 \pm 0.007$ & $8.263 \pm 0.018$ & $12.771 \pm 0.058$ \\
$1.0$ & $3.014 \pm 0.003$ & $3.604 \pm 0.003$ & $4.475 \pm 0.004$ & $5.866 \pm 0.007$ & $8.348 \pm 0.018$ & $12.906 \pm 0.058$ \\
$1.2$ & $3.039 \pm 0.003$ & $3.628 \pm 0.003$ & $4.498 \pm 0.004$ & $5.882 \pm 0.007$ & $8.335 \pm 0.018$ & $12.949 \pm 0.059$ \\
\hline
\end{tabular}
\caption{$\Sigma(z)$ with uncertainties for linear oscillations.}
\label{tab:LinOsc_sigma_z_grid}
\end{table}

\newpage
\section{Full Residual Grids and Simulation Uncertainty}
\label{full_results}
\begin{figure}[h!]
    \centering
    \includegraphics[scale=0.3]{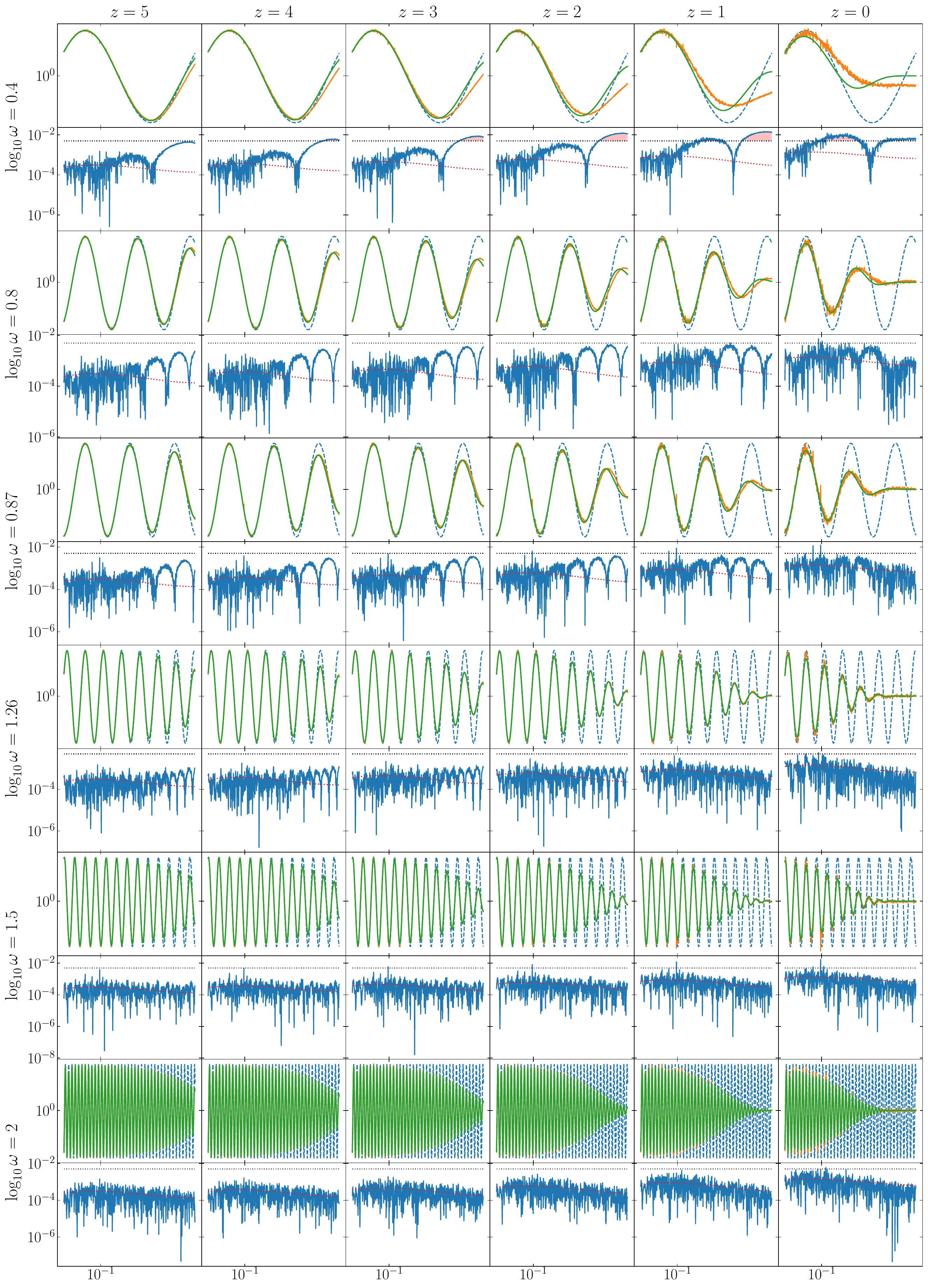}
    \caption{Residuals (blue, bottom row) between the model fit (green, top row) and simulated
    non-linear matter power spectrum (orange, top row) compared to the polynomial fit to the simulation standard deviation (red dashed line, bottom row) as a function of redshift (left to right) and frequency (top to bottom)}
    \label{fig:FullLogResults}
\end{figure}
\begin{figure}[h]
    \centering
    \includegraphics[scale=0.3]{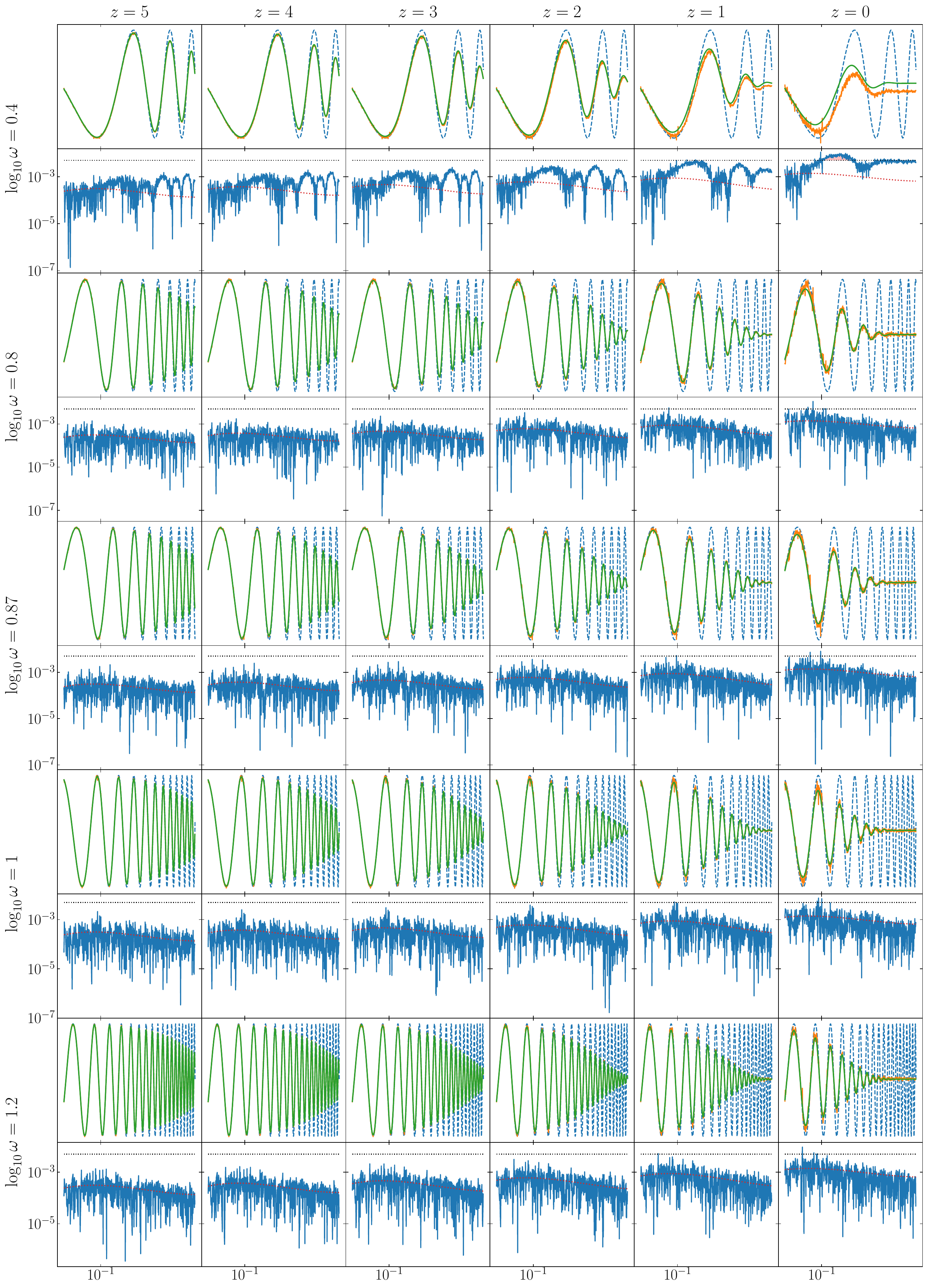}
    \caption{Residuals (blue, bottom row) between the model fit (green, top row) and simulated
    non-linear matter power spectrum (orange, top row) compared to the polynomial fit to the simulation standard deviation (red dashed line, bottom row) as a function of redshift (left to right) and frequency (top to bottom)}
    \label{fig:FullLinearResults}
\end{figure}
\newpage

\end{appendix}
\end{document}